\documentclass[10pt]{article}
\usepackage[margin=1in]{geometry} 
\usepackage{amsmath,amsthm,amssymb, color}
\usepackage{enumerate} 
\usepackage[colorlinks=true, linkcolor=blue]{hyperref}
\usepackage{color}
\usepackage{multirow}
\usepackage{enumitem}
\usepackage{mathtools}
\usepackage{braket}
\usepackage{booktabs}
\usepackage{tabularx}
\usepackage{threeparttable}
\usepackage{tikz}
\usetikzlibrary{positioning}
\usepackage{caption}
\usepackage{subcaption}
\usepackage[compat=0.4]{yquant}
\usetikzlibrary{quotes}
\usepackage{graphicx} 
\RequirePackage{algorithm}
\RequirePackage{algorithmic}

\usepackage{dsfont}
\usepackage{amsfonts}
\usepackage{tablefootnote}
\usepackage{adjustbox}
\usepackage{float}
\usepackage[capitalize]{cleveref}
\usepackage{bbold}
\usepackage{tcolorbox}
\usepackage{thmtools}
\usepackage{thm-restate}
\usepackage{authblk}

\allowdisplaybreaks

\theoremstyle{plain}
\newtheorem{theorem}{Theorem}[section]
\newtheorem{proposition}[theorem]{Proposition}
\newtheorem{lemma}[theorem]{Lemma}

\theoremstyle{definition}

\newtheorem{assumption}[theorem]{Assumption}
\theoremstyle{remark}
\newtheorem{remark}[theorem]{Remark}
\theoremstyle{plain}
\newtheorem{question}{Question}

\newcommand{\x}{\mathbf{x}}
\newcommand{\y}{\mathbf{y}}

\newcommand{\q}{\mathbf{q}}
\newcommand{\p}{\mathbf{p}}
\newcommand{\g}{\mathbf{g}}
\newcommand{\uu}{\mathbf{u}}
\newcommand{\eps}{\boldsymbol{\epsilon}}

\newcommand{\KL}{\textnormal{KL}}
\newcommand{\FI}{\textnormal{FI}}
\newcommand{\TV}{\textnormal{TV}}
\newcommand{\Ocal}{\mathcal{O}}
\newcommand{\W}{\textnormal{W}}
\newcommand{\bbe}{\mathbb{E}}
\newcommand{\algname}[1]{\textup{\texttt{#1}}}

\crefname{assumption}{Assumption}{Assumptions}
\crefname{theorem}{Theorem}{Theorems}
\crefname{lemma}{Lemma}{Lemmas}
\crefname{proposition}{Proposition}{Propositions}

\title{Quantum Speedups for Sampling and Non-convex Optimization with Stochastic Oracles}
\author[1]{Guneykan Ozgul}
\author[2]{Xiantao Li}
\author[1]{Mehrdad Mahdavi}
\author[1]{Chunhao Wang}

\affil[1]{Department of Computer Science and Engineering, Pennsylvania State University, University Park, PA, USA}
\affil[2]{Department of Mathematics, Pennsylvania State University, University Park, PA, USA}
\affil[ ]{\texttt{guneykanozgul@gmail.com}, \texttt{xiantao.li@psu.edu}, \texttt{mzm616@psu.edu}, \texttt{cwang@psu.edu}}

\date{}

\begin{document}

\maketitle
\begin{abstract}
We present quantum speedups for sampling from distributions of the form $\pi\propto e^{-f}$ on $\mathbb{R}^d$. We consider two stochastic oracle models: a stochastic gradient oracle, where $f=\frac{1}{n}\sum_{i=1}^n f_i $ and component gradients $\{\nabla f_i\}_{i \in [n]}$ are available, and a stochastic evaluation oracle, where only noisy values of $f$ are available. Our framework accelerates classical stochastic Langevin Monte Carlo (LMC) and Hamiltonian Monte Carlo (HMC) algorithms by replacing stochastic gradient estimators with variance-controlled quantum mean estimation and gradient estimation subroutines. Unlike quantum walk based approaches, our algorithms do not require reversibility or exact gradients, and they preserve the structure of the underlying Markov chain. In the finite-sum setting, quantum mean estimation combined with classical variance-reduction techniques improves the stochastic gradient-query complexity for the approximate sampling task. In the stochastic zeroth-order setting, we develop gradient estimators robust to noisy function evaluations, yielding improved evaluation complexity for LMC and HMC. These results apply to strongly log-concave and/or non-log-concave distributions satisfying a log-Sobolev inequality, with convergence guarantees in Wasserstein distance and Kullback--Leibler divergence. We also show that faster sampling methods lead to quantum speedups for optimization, including for non-smooth and approximately convex objectives.
\end{abstract}
\clearpage
\tableofcontents
\clearpage

\section{Introduction}
\label{introduction}
\subsection{Motivation}
Efficient sampling from complex distributions is a fundamental problem in many scientific and engineering disciplines and it is becoming increasingly important as modern applications deal with high-dimensional data and complex probabilistic models.
In machine learning, sampling plays an important role in Bayesian inference, as it facilitates posterior estimation and uncertainty quantification~\cite{10.5555/3104482.3104568,pmlr-v37-wangg15,durmus2018highdimensionalbayesianinferenceunadjusted,roy2021stochasticzerothorderdiscretizationslangevin}. In non-convex optimization, sampling allows for the exploration of complex energy landscapes and helps avoid local minima, facilitating progress in tasks such as resource allocation, scheduling, and hyperparameter tuning~\cite{pmlr-v65-zhang17b,JMLR:v21:19-327}. It is also widely used in other domains of science, such as statistical mechanics~\cite{chandler1987introduction,dfrenkel96:mc}, and convex geometry \cite{LOVASZ2006392, doi:10.1137/15M1054250}. 

Given a potential function $f:\mathbb{R}^d\to \mathbb{R}$, we consider the problem of approximately sampling from \emph{Gibbs-Boltzmann distribution} $\pi\in L^1(\mathbb{R}^d, \mathrm{d}\x)$ of the form
\begin{equation}
\label{eq:Gibbs-Boltzmann}
\pi(\x) = \frac{e^{-f(\x)}}{\int e^{-f(\x)}\mathrm{d}\x}.
\end{equation}
Solving this problem is in general difficult as it encodes NP-hard problems, and efficient sampling is only possible when $f$ is highly structured. Given the challenging nature of the problem, it is natural to ask whether quantum computers can provide any speedup for the sampling task.
The state-of-the-art classical algorithms are based on Markov chains, and this is a domain where quantum algorithms have been studied extensively \cite{1366222,PhysRevA.78.042336, 10.5555/3370245.3370246}. 
In the continuous setting, Childs et al.~\cite{NEURIPS2022_933e9533} initiated the study of quantum algorithms for sampling from log-concave distributions, assuming that \(f\) is strongly convex and that \(\nabla f\) can be accessed to very high precision. Their construction uses the quantum-walk framework of~\cite{1366222} together with fixed-point amplitude amplification similar to~\cite{PhysRevA.78.042336}. Although this approach improves the best classical dependence on certain parameters, the speedup does not directly apply in the stochastic setting because the relevant Markov chains are not \emph{reversible}. Ozgul et al.~\cite{pmlr-v235-ozgul24a} later addressed this limitation by analyzing a reversible proxy chain; as a result, the speedup is obtained relative to a suboptimal version of the best classical sampler. 

Moreover, existing quantum results in this line primarily guarantee convergence in total variation distance. Obtaining quantum speedups with guarantees in Wasserstein distance or Kullback-Leibler divergence, metrics that are standard in classical sampling theory, remains largely open. Motivated by these challenges, we ask the following question:
\begin{question} 
\label{question1} 
Can quantum computers improve the oracle complexity of sampling algorithms such that \begin{enumerate} 
\item the algorithm can use stochastic gradient and/or noisy evaluation oracles, 
\item it can be based on fast (potentially nonreversible) Markov chains rather than a reversibilized surrogate, and 
\item it provides guarantees in Wasserstein distance and Kullback--Leibler divergence? \end{enumerate}
\end{question}

We answer Question~\ref{question1} affirmatively by introducing a framework that replaces stochastic gradient computation with variance-controlled quantum subroutines, yielding provable oracle-complexity improvements while keeping the underlying nonreversible Markov chain unchanged. Conceptually, our contribution is not a new sampler, but a principled way to accelerate existing stochastic samplers by exploiting variance structure (via quantum mean/gradient estimation) rather than relying on restrictive quantum-walk techniques. 

Another motivation for this work is to investigate whether quantum speedups for certain optimization problems can be obtained via our sampling algorithms. Sampling and optimization are tightly connected: sampling from a Gibbs distribution with potential $\beta f$ at large inverse temperature $\beta$ concentrates around minimizers of $f$. Although using sampling as a generic optimization method is not always optimal, it can be effective in regimes where gradient information is noisy or the objective is non-smooth/non-convex. To this end, we ask the following question:

\begin{question}
\label{question2}
    Can we use the quantum sampling algorithms to recover the existing quantum speedups for certain optimization problems and/or identify new classes of optimization problems that can be accelerated through sampling?
\end{question}

We also answer Question~\ref{question2} affirmatively by designing optimization algorithms based on our quantum samplers. In the finite-sum setting, we recover the quantum speedup of~\cite{pmlr-v235-zhang24bz}; in the zeroth-order setting, we improve some parameter dependencies for non-smooth optimization problems studied in~\cite{10.5555/3600270.3600498,chakrabarti2025speedupsconvexoptimizationquantum}. 
While most of the resulting improvements are sub-quadratic, the point of our work is not to focus on the degree of the speedup for these particular problems. More broadly, we highlight that viewing optimization through the lens of sampling can be a useful organizing principle for identifying future quantum speedups. The key point is that sampling becomes competitive with standard optimization methods in regimes where (i) gradients are unavailable or prohibitively expensive, (ii) the objective is non-smooth (the gradients are not Lipschitz continuous), or (iii) the landscape is non-convex and escaping shallow traps is important. In these regimes, a low-temperature Gibbs distribution can serve as a powerful tool to approach these problems.

This perspective is also useful from a quantum-algorithmic standpoint for two related reasons. First, many standard first-order optimization routines (e.g., gradient descent) do not admit meaningful generic quantum speedups in the usual black-box models. Second, the sampling problem offers a more natural ground for quantum algorithms compared to deterministic iterative procedures. Moreover, in non-convex settings, which are a primary target for quantum acceleration, sampling-based methods are shown to be particularly effective~\cite{Ma_2019}. 

We use Langevin Monte Carlo (LMC) and Hamiltonian Monte Carlo (HMC) as base samplers. Both are sensitive to gradient noise: smaller variance allows larger steps or fewer iterations, but producing a lower-variance gradient estimator costs more oracle queries. Detailed descriptions of their stochastic-gradient variants are given alongside the corresponding results in~\cref{sec:lmc-overview,sec:hmc-overview}.

In the finite-sum setting, a stochastic gradient can be obtained by sampling an index $i\in [n]$ and using $\nabla f_i(\x)$ as an approximation to $\nabla f(\x)$. 
This estimator is unbiased, but because of the large variance, the error due to stochasticity may be too large for sampling unless the step size is made very small.
In zeroth-order settings, stochastic gradients are often built from finite differences or Gaussian smoothing. 
Such estimators have an additional dimension-dependent cost because a classical method must probe enough directions to recover a $d$-dimensional gradient.
The quantum algorithms in this paper improve these two bottlenecks: the finite-sum algorithms reduce the cost of estimating $\nabla f = \frac{1}{n} \sum_{i=1}^n \nabla f_i$ using queries to $\{\nabla f_i\}_{i\in [n]}$, while the zeroth-order algorithms reduce the cost of estimating $\nabla f$ using queries to an oracle that returns $f$ with some noise.

\subsection{Quantum Oracle Model}
Here, we describe our quantum input model and assumptions.
\medskip

\noindent\textbf{Stochastic gradient oracle.} In this setting, we assume $f(\mathbf{x}) = \frac{1}{n}\sum_{i=1}^n f_i(\mathbf{x})$, and we assume access to the gradients of each component, i.e., $\nabla f_i(x)$ for any $i\in [n]$. 
  This is a relevant setting in statistics and machine learning problems, where access to the function $f$ is provided only through sample data~\cite{10.5555/3104482.3104568}. 
  For each $i\in[n]$, we consider the quantum oracle access of the form
\begin{equation}
    \label{eq:oracle}
    O_{\nabla f}\ket{\x}\ket{i}\ket{0}\mapsto \ket{\x}\ket{i}\ket{\nabla f_i(\x)}.
\end{equation}

\medskip

\noindent\textbf{Stochastic evaluation oracle.} In this setting, we assume that we have access only to noisy function values $f(\x;\xi)$, where $\xi$ is the random seed that characterizes the noise.
  This setting is more relevant in contexts where the function value $f$ is replaced by a statistical estimate, perhaps obtained through a numerical simulation~\cite{roy2021stochasticzerothorderdiscretizationslangevin}.
  The function is accessed through the binary oracle below.
\begin{equation}
\label{eq:binary-oracle}
    O_{f}\ket{\x}\ket{\xi}\ket{0}\mapsto \ket{\x}\ket{\xi}\ket{f(\x;\xi)}.
\end{equation}
Since this model only uses function values, we use the terms ``zeroth-order oracle'' and ``evaluation oracle'' interchangeably. We note that the evaluation oracle requires coherent access to the random seed \(\xi\) and that the same seed can be queried at multiple points. 
Although this model is stronger than a standard classical noisy oracle, we show that it naturally captures non-smooth optimization settings. This assumption is also explicit in the classical baseline work~\cite{roy2021stochasticzerothorderdiscretizationslangevin}.

Both oracles can be implemented by making the corresponding classical oracles reversible with additional registers, so their query cost is comparable to the classical setting. 
As our study spans multiple settings under various assumptions, we state the precise noise assumptions separately for each setting.

\subsection{Main Contributions}
We summarize our technical contributions as follows. 

\medskip
\noindent\textbf{Speedups for sampling in the stochastic gradient oracle setting}
Using quantum mean estimation together with classical variance-reduction techniques such as stochastic variance-reduced gradients (SVRG) and control variates (CV), we obtain improved query complexities for sampling from both convex and non-convex potentials in~\cref{sec:finite-sum}. The main challenge is that these algorithms periodically compute the full gradient to control the variance of the stochastic gradients, and this full-gradient cost can dominate the improvement obtained from quantum mean estimation. To address this, we analyze the variance of the stochastic gradients and show that one can choose the step size and target variance online so that full-gradient computations are needed less frequently when quantum variance reduction is used. We also use these samplers to construct optimization algorithms for strongly convex functions in~\cref{sec:finite-sum-optimization}.

\medskip
\noindent \textbf{Quantum speedups for gradient estimation via stochastic evaluation oracle}
The challenge in this setting is that existing gradient estimation algorithms assume that the evaluation oracle is almost exact. 
However, in the setting we consider, the error between the estimated function value and the true function value can be unbounded. 
To address this, we develop quantum gradient estimation algorithms under several smoothness assumptions in~\cref{sec:zeroth-order-gradient}.
 When the potential is smooth, our estimator reduces the number of evaluation queries from \(\Tilde{\Ocal}(d^2\sigma^2/\epsilon^2)\) to \(\Tilde{\Ocal}(d\sigma/\epsilon)\) for computing an \(\epsilon\)-accurate gradient estimate (\cref{thm:zeroth-order-gradient1}), where \(\sigma^2\) is the noise variance in~\cref{ass:bounded-variance-zeroth-order}. When the stochastic functions are also smooth with high probability, we improve the dimension dependence by an additional factor of \(d^{1/2}\) (\cref{thm:smooth-gradient}).
Our gradient estimation algorithms could be useful as independent tools, especially in zeroth-order stochastic optimization. 

\medskip
\noindent \textbf{Speedups for stochastic zeroth-order sampling.} In~\cref{sec:sampling-via-zeroth-order}, we combine our quantum gradient estimation algorithm with HMC and LMC and analyze the convergence of the resulting samplers. We prove that, under the same assumptions, these hybrid algorithms use fewer queries to the evaluation oracle than the best-known classical samplers (\cref{thm:qzhmc,thm:qzlmc}).

\medskip
\noindent \textbf{Application to non-smooth non-convex optimization.} In~\cref{sec:application}, we extend our quantum sampling methods to optimize non-convex functions with specific structural properties without the smoothness assumption. In particular, we show that Lipschitz continuous and approximately convex functions can be optimized with fewer stochastic-evaluation queries than the best-known classical algorithms in terms of dimension dependence (\cref{thm:optimization}).
\medskip

To help the readers navigate, we summarize the problems, oracles, assumptions, and results in \cref{table:allresults}.

\begin{table}[t]
\centering
\caption{Overview of the main results. The table separates assumptions on the oracle/noise model from assumptions on the target distribution or optimization objective.}
\label{table:allresults}
\scriptsize
\setlength{\tabcolsep}{3pt}
\renewcommand{\arraystretch}{1.18}
\begin{tabularx}{\textwidth}{@{}
>{\raggedright\arraybackslash}p{1.9cm}
>{\raggedright\arraybackslash}p{1.75cm}
>{\raggedright\arraybackslash}X
>{\raggedright\arraybackslash}X
>{\raggedright\arraybackslash}p{2.2cm}@{}}
\toprule
\textbf{Problem} & \textbf{Oracle} & \textbf{Oracle/noise assumptions} & \textbf{Target or objective assumptions} & \textbf{Results} \\
\midrule
Sampling & Stochastic\ gradient & Lipschitz stochastic gradients & Strong convexity\ and Smoothness& \cref{thm:qhmc-svrg-main,thm:qhmc-cv-main} \\
\addlinespace[0.25em]
Sampling & Stochastic\ gradient & Lipschitz stochastic gradients & Log-Sobolev inequality\ and Smoothness & \cref{thm:qlmc} \\
\addlinespace[0.25em]
Sampling & Stochastic\ evaluation & Bounded noise variance & Strong convexity\ and Smoothness& \cref{thm:qzhmc} \\
\addlinespace[0.25em]
Sampling & Stochastic\ evaluation &  Bounded noise variance & Log-Sobolev inequality\ and Smoothness & \cref{thm:qzlmc} \\
\addlinespace[0.25em]
Optimization & Stochastic\ evaluation & Uniform error bound & Approximate convexity\ and Lipschitz continuity & \cref{thm:optimization} \\
\bottomrule
\end{tabularx}
\end{table}

\subsection{Related Work}
Non-asymptotic convergence rates for stochastic LMC and HMC have been analyzed extensively by~\cite{pmlr-v65-raginsky17a,NEURIPS2018_9c19a2aa,pmlr-v161-zou21a,pmlr-v195-das23b} and~\cite{pmlr-v32-cheni14,pmlr-v139-zou21b}, respectively. In the finite-sum setting, more sophisticated variance-reduction techniques, such as SVRG, SAGA, SARAH, and control variates (CV)~\cite{NIPS2013_ac1dd209,NIPS2014_ede7e2b6,pmlr-v70-nguyen17b,10.1007/s11222-018-9826-2}, have been used to reduce stochastic-gradient variance by using gradient information from previous iterations.
Although these methods were originally introduced in the context of optimization, successive works have applied these methods to improve sampling efficiency via LMC \cite{NIPS2016_9b698eb3,pmlr-v80-chatterji18a,10.1007/s11222-018-9826-2,NEURIPS2022_78e839f9} and HMC \cite{NEURIPS2019_c3535feb,pmlr-v139-zou21b}. In particular, \cite{pmlr-v139-zou21b} has incorporated various variance reduction techniques into SG-HMC and analyzed convergence in Wasserstein distance for smooth and strongly convex potentials. In the non-log-concave setting, \cite{NEURIPS2022_78e839f9} has analyzed the convergence of SVRG-LMC and SARAH-LMC for target distributions that satisfy the log-Sobolev inequality and applied their results to optimize structured non-convex objectives.

In quantum sampling, most existing results use quantum walks, which can speed up certain Markov Chain Monte Carlo (MCMC) methods by improving the mixing time of the underlying Markov chain~\cite{1366222,somma2007quantumsimulatedannealing,PhysRevLett.101.130504,PhysRevA.78.042336,10.1145/3588579}. These methods have been used in several domains to reduce the computational time of sampling-related tasks~\cite{10.1145/1250790.1250874,10.5555/3370245.3370246,NEURIPS2022_933e9533,10.1145/3588579,10.5555/3600270.3600498,chakrabarti2024generalizedshortpathalgorithms}. For continuous distributions, \cite{NEURIPS2022_933e9533, pmlr-v235-ozgul24a} used the quantum-walk framework to reduce the number of gradient queries needed to approximately sample from strongly log-concave and non-log-concave distributions.

In the context of optimization, quantum algorithms such as multi-dimensional quantum mean estimation~\cite{Cornelissen_2022} and quantum gradient estimation~\cite{Jordan_2005,Gily_n_2019} have shown promise in reducing the computational cost associated with gradient-based methods \cite{vanApeldoorn2020convexoptimization,Chakrabarti2020quantumalgorithms,NEURIPS2023_6ed9931d,pmlr-v235-zhang24bz,liu2024quantum}. These techniques are particularly well-suited for addressing challenges in large-scale and noisy settings, as they can provide more accurate gradient estimates with fewer queries. However, these methods have not been considered for sampling tasks, and this paper focuses on integrating these quantum techniques to enhance the efficiency of stochastic gradient-based samplers and alleviate the computational burden inherent in classical methods.

\subsection{Notation and Glossary}
Bold symbols, such as $\x$ and $\y$, are used to represent vectors, with $\|\cdot\|$ indicating the Euclidean or operator norm depending on the context. Given two scalars $a$ and $b$, we use $a \wedge b$ to denote $\min\{a, b\}$ and use $a\vee b$ to denote $\max\{a, b\}$. We use $\mathcal{B}_d(c,r)$ to denote the $d$-dimensional ball centered at $c$ with radius $r$ and $G_d^l(c)$ to denote the $d$-dimensional grid centered at point $c$ with side length $l$. We occasionally use $G_d^l$ when the center of the grid is clear from the context. The notation $\Tilde{O}$ is used to suppress the polylogarithmic dependencies on $d, \epsilon, L,\mu$ and $\alpha$ that will be defined later in the text.

We use several metrics to compare probability distributions over a state space \( \mathcal{X} \). Let \( \pi \) and \( \mu \) be two probability distributions on \( \mathcal{X} \). The \( p \)-Wasserstein distance between \( \pi \) and \( \mu \) is defined as 
$\text{W}_p(\pi, \mu) = \left( \inf_{\gamma \in \Gamma(\pi, \mu)} \mathbb{E}_{(\x, \y) \sim \gamma} \|\x - \y\|^p \right)^{1/p}$ where \( \Gamma(\pi, \mu) \) is the set of all joint distributions \( \gamma(\x, \y) \) whose marginals are \( \pi \) and \( \mu \).
The KL divergence of $\pi$ with respect to $\mu$ is defined as
$\KL(\pi \| \mu) = \int_{\mathcal{X}} \mathrm{d}\x\pi(\x) \log \left( \frac{\pi(\x)}{\mu(\x)} \right)
$ and the relative Fisher information is
$ 
\FI(\pi \| \mu) = \int_{\mathcal{X}} \mathrm{d}\x\pi(\x) \left\|\nabla \log \left( \frac{\pi(\x)}{\mu(\x)} \right)\right\|^2
$.
The total variation distance is defined as 
$
    \TV(\pi, \mu) = \sup_{A \subseteq \mathcal{X}} |\pi(A) - \mu(A)| = \frac{1}{2} \int_{\mathcal{X}} \mathrm{d}\x|\pi(\x) - \mu(\x)|
$.

\subsection{Organization of the paper}
In~\cref{sec:finite-sum}, we present our quantum sampling algorithms for stochastic gradient oracles. In~\cref{sec:zeroth-order-gradient}, we introduce the stochastic zeroth-order gradient estimation algorithm that serves as the main technical tool for the sampling results in~\cref{sec:sampling-via-zeroth-order}. Finally, in~\cref{sec:application}, we show how these sampling results can be used to speed up non-convex optimization.

\section{Quantum Speedups for Finite-Sum Sampling and Optimization via Gradient Oracle}
\label{sec:finite-sum}
Throughout this section, we consider \(f=\frac1n\sum_{i=1}^n f_i\), and use the quantum stochastic-gradient oracle in~\cref{eq:oracle}. Our goal is to sample approximately from \(\pi\) using as few oracle queries as possible. We first introduce stochastic-gradient estimators that combine unbiased quantum mean estimation with classical variance-reduction frameworks, yielding improved query complexity for both LMC and HMC.
Classically, a stochastic gradient at \(\x_t\) is often formed by subsampling a mini-batch \(B\subseteq [n]\) and setting \(\g_t=|B|^{-1}\sum_{i\in B}\nabla f_i(\x_t)\). Then \(\mathbb{E}_B[\g_t]=\nabla f(\x_t)\), so \(\g_t\) is an unbiased estimator. If LMC or HMC takes \(T\) iterations to converge, the resulting query complexity is \(\Ocal(|B|T)\), and there is a tradeoff between the batch size \(|B|\) and the number of iterations \(T\).

To decouple these quantities, variance-reduction methods such as SVRG periodically compute a full gradient \(\nabla f(\tilde{\x})\) and use it to correct later stochastic gradients. Control variates (CV) use a similar correction based on the initial gradient \(\nabla f(\x_0)\). Our quantum LMC algorithms preserve this structure, but replace mini-batch averaging with unbiased quantum mean estimation; see \cref{algo:qsvrg-lmc,algo:qsvrg-cv}.

\subsection{Quantum Mean Estimation and Variance Reduction}
Quantum mean estimation estimates the mean of a $d$-dimensional random variable $X$ to accuracy $\epsilon$ using $\Tilde{\Ocal}(d^{1/2}/\epsilon)$ queries, which is a quadratic improvement in $\epsilon$ over classical algorithms~\cite{Cornelissen_2022}. Although the original quantum mean estimation algorithm is biased, the method of~\cite{NEURIPS2023_6ed9931d} is unbiased. Specifically, for a multidimensional variable with mean $\mu$ and variance $\sigma^2$, it outputs an estimate $\hat{\mu}$ such that $\mathbb{E}[\hat{\mu}]=\mu$ and $\mathbb{E}[\|\hat{\mu}-\mu\|^2]\leq \hat{\sigma}^2$ using $\Tilde{\Ocal}(d^{1/2}\sigma/\hat{\sigma})$ queries.

\begin{lemma}[Unbiased quantum mean estimation~\cite{NEURIPS2023_6ed9931d}]
\label{lem:var-reduction}
Let $X$ be a $d$-dimensional random variable with $\mathrm{Var}[X]\leq\sigma^2$, and suppose a quantum sampling oracle for $X$ is available. For any target variance $\hat{\sigma}^2$, there is a procedure that outputs an unbiased estimate of $\bbe X$ with variance at most $\hat{\sigma}^2$ using $\widetilde{\Ocal}(d^{1/2}\sigma/\hat{\sigma})$ oracle queries.
\end{lemma}

\begin{assumption}[Lipschitz Stochastic Gradients]
There exists a positive constant $L$ such that for all $\x, \y \in \mathbb{R}^d$ and all
functions $f_i$, $i = 1, ..., n$, it holds that
\label{assumption:component-smoothness}
    \begin{equation}
        \|\nabla f_i(\x)- \nabla f_i(\y) \| \leq L \|\x - \y\|.
    \end{equation}
\end{assumption}
This is a standard assumption in the optimization and sampling literature~\cite{NEURIPS2019_65a99bb7,Ma_2019}.

\subsection{Sampling under Log-Sobolev Inequality via Langevin Monte Carlo}
\label{sec:lmc-overview}
Langevin Monte Carlo (LMC), also called the unadjusted Langevin algorithm, discretizes the continuous-time Langevin diffusion
\begin{equation}
\label{eq:langevin-sde}
\mathrm{d}\x_t=-\nabla f(\x_t)\mathrm{d}t+\sqrt{2}\mathrm{d}\mathbf{B}(t),
\end{equation}
where $\mathbf{B}(t)$ is standard Brownian motion. The target distribution $\pi$ is stationary for this diffusion. Its Euler--Maruyama discretization with a stochastic gradient $\g_k$ is
\begin{equation}
\label{eq:LMC}
\x_{k+1}=\x_k-\eta_k\g_k+\sqrt{2\eta_k}\eps_k,
\end{equation}
where $\eps_k\sim\mathcal{N}(0,I_d)$. Stochastic-gradient LMC reduces the per-step cost when a full gradient is expensive or unavailable, but the variance of $\g_k$ introduces additional sampling error.

\begin{algorithm}[ht]
\begin{algorithmic}[1]
\caption{$\algname{SG-LMC}$}
\label{algo:LMC}
\INPUT Stochastic gradient routine $\g(\cdot)$, initial point $\x_0$, step sizes $\{\eta_k\}_{k=0}^{K-1}$, number of steps $K$.
\OUTPUT Approximate sample from $\pi\propto e^{-f(\x)}$.
\FOR{$k=0,\ldots,K-1$}
\STATE Sample $\eps_k\sim\mathcal{N}(0,I_d)$.
\STATE Set $\x_{k+1}=\x_k-\eta_k\g(\x_k)+\sqrt{2\eta_k}\eps_k$.
\ENDFOR
\STATE \textbf{Return} $\x_K$.
\end{algorithmic}
\end{algorithm}

We use the SVRG-LMC algorithm of~\cite{NEURIPS2022_78e839f9} as the base sampler and replace its stochastic-gradient estimator with the quantum estimator in \(\algname{QSVRG-LMC}\); see \cref{algo:qsvrg-lmc}. In this section, we assume that the target distribution satisfies a log-Sobolev inequality.

\begin{assumption}[Log-Sobolev Inequality]
        \label{assumption:LSI}
         We say that $\pi$ satisfies the log-Sobolev inequality with constant $\alpha$ if for all $\rho$, it holds that
       \begin{align}
       \label{eq:LSI}
           \KL(\rho||\pi)\leq \frac{1}{2\alpha}\FI(\rho||\pi).
       \end{align}
\end{assumption}

The log-Sobolev inequality is a sampling analog of the Polyak-Łojasiewicz (PL) condition used in optimization~\cite{chewi2024ballisticlimitlogsobolevconstant}.
It is standard in the non-log-concave sampling literature~\cite{NEURIPS2019_65a99bb7,Ma_2019,pmlr-v178-chewi22a,NEURIPS2022_78e839f9}. Moreover, if \(f\) is \(\mu\)-strongly convex, then \(\pi\) satisfies an LSI with constant $\mu$.  We also note that this assumption is weaker than the dissipative gradient condition~\cite{pmlr-v65-raginsky17a,NEURIPS2019_c3535feb} which is commonly used in non-log-concave sampling. 

\begin{algorithm}[ht]
\begin{algorithmic}[1]
\caption{$\algname{QSVRG-LMC}$}
\label{algo:qsvrg-lmc}
\INPUT Stochastic gradient oracle $O_{\nabla f}$, initial point $\x_0$, number of iterations $K$, step size $\eta$, smoothness constant $L$, variance scale factor $b$, epoch length $m$.
\OUTPUT Final iterate $\x_K$.
\STATE Set $\tilde{\x}=\x_0$ and $\tilde{\g}=\nabla f(\tilde{\x})$.
\FOR{$k=0,\ldots,K-1$}
    \IF{$k \bmod m = 0$}
        \STATE Set $\tilde{\x}=\x_k$ and $\tilde{\g}=\nabla f(\tilde{\x})$.
        \STATE Set $\g_k=\tilde{\g}$.
    \ELSE
        \STATE Define the quantum sampling oracle $O_{\mathrm{SVRG}}^{\x_k,\tilde{\x}}$ by
        \[
        \ket{0}\ket{0}\mapsto \frac{1}{\sqrt{n}}\sum_{i=1}^n
        \ket{\nabla f_i(\x_k)-\nabla f_i(\tilde{\x})+\tilde{\g}}\ket{i}.
        \]
        \STATE Set $\hat{\sigma}_k^2=L^2\|\x_k-\tilde{\x}\|^2/b^2$.
        \STATE Set $\g_k=\algname{QuantumMeanEstimation}(O_{\mathrm{SVRG}}^{\x_k,\tilde{\x}},\hat{\sigma}_k^2)$.
    \ENDIF
    \STATE Sample $\eps_k\sim\mathcal{N}(0,I_d)$.
    \STATE Set $\x_{k+1}=\x_k-\eta\g_k+\sqrt{2\eta}\eps_k$.
\ENDFOR
\STATE \textbf{Return} $\x_K$.
\end{algorithmic}
\end{algorithm}

\begin{algorithm}[ht]
\begin{algorithmic}[1]
\caption{$\algname{QCV-LMC}$}
\label{algo:qsvrg-cv}
\INPUT Stochastic gradient oracle $O_{\nabla f}$, initial point $\x_0$, number of iterations $K$, step size $\eta$, smoothness constant $L$, variance scale factor $b$.
\OUTPUT Final iterate $\x_K$.
\STATE Set $\g^{(0)}=\nabla f(\x_0)$.
\FOR{$k=0,\ldots,K-1$}
    \STATE Define the quantum sampling oracle $O_{\mathrm{CV}}^{\x_k,\x_0}$ by
    \[
    \ket{0}\ket{0}\mapsto \frac{1}{\sqrt{n}}\sum_{i=1}^n
    \ket{\nabla f_i(\x_k)-\nabla f_i(\x_0)+\g^{(0)}}\ket{i}.
    \]
    \STATE Set $\hat{\sigma}_k^2=L^2\|\x_k-\x_0\|^2/b^2$.
    \STATE Set $\g_k=\algname{QuantumMeanEstimation}(O_{\mathrm{CV}}^{\x_k,\x_0},\hat{\sigma}_k^2)$.
    \STATE Sample $\eps_k\sim\mathcal{N}(0,I_d)$.
    \STATE Set $\x_{k+1}=\x_k-\eta\g_k+\sqrt{2\eta}\eps_k$.
\ENDFOR
\STATE \textbf{Return} $\x_K$.
\end{algorithmic}
\end{algorithm}

For \(\algname{QSVRG}\), the epoch length \(m\) determines how often the full gradient is computed. The parameter \(b\) is the quantum analogue of the batch size and is chosen analytically. In \(\algname{QCV}\), the stochastic gradient is corrected using the initial iterate, as in classical control variates. We also highlight that quantum mean estimation requires a target variance level as input, unlike classical subsampling. Although the optimal target variance changes along the trajectory and depends on quantities that are difficult to compute, \(\algname{QSVRG}\) uses \(L^2\|\x_k-\tilde{\x}\|^2/b^2\), where \(\tilde{\x}\) is the most recent point at which the full gradient was computed~(\cref{algo:qsvrg-lmc}). Similarly, \(\algname{QCV}\) uses \(L^2\|\x_k-\x_0\|^2/b^2\), where \(\x_0\) is the initial point~(\cref{algo:qsvrg-cv}). Note that we only use \(\algname{QCV}\) for HMC, but \cref{algo:qsvrg-cv} uses an LMC update for ease of exposition. Replacing that update with the HMC proposal in~\cref{algo:HMC} implements \(\algname{QCV-HMC}\).

Although the algorithm is simple, proving the speedup is more delicate because the cost of full-gradient computations must also be reduced. We show that quantum mean estimation enables less frequent full-gradient computations through improved variance reduction. Our analysis quantifies stochastic-gradient variance along the algorithm trajectory and balances the frequency of full-gradient computations with the target variance used in quantum mean estimation. This yields improved complexity bounds in both strongly convex and non-convex settings. We first present the LMC result without assuming convexity; later, we consider HMC with similar stochastic-gradient estimators to further improve some parameter dependencies.

We first control the stochastic-gradient variance along the trajectory.
\begin{restatable}[$\algname{QSVRG-LMC}$ Variance Lemma]{lemma}{svrgvariance}
\label{lem:svrgvariance}
Let $k'<k$ be the last iteration where the full gradient is computed in \algname{QSVRG-LMC} and $\sigma_k^2 = \mathbb{E}\|\g_k - \nabla f(\x_k) \|^2$. Then, for $\eta^2 \leq \frac{1}{6L^2m^2}$,
\begin{align}
    \sigma_{k'+l}^2\leq \frac{16L^4 \eta^2 }{\alpha}\sum_{r=1}^{l} \KL(\mu_{k'+r-1}||\pi) + \frac{8\eta d m L^2}{b^2}.
\end{align}
\end{restatable}
The proof is given in \cref{sec:proof-qsvrg-lmc}. This variance estimate yields the following convergence theorem.
\begin{restatable}[Convergence theorem for \algname{QSVRG-LMC}]{theorem}{qsvrgconvergence}
\label{thm:qsvrgconvergence}
Assume that $m\leq b^2$. Then, for $\eta \leq \frac{\alpha^2}{24 L^2 m}$, the iterates in \algname{QSVRG-LMC} satisfy
\begin{align}
     \KL(\mu_{k}||\pi)\leq e^{-\alpha \eta k}\KL(\mu_0\|\pi)+\frac{64m \eta dL^2}{\alpha b^2}+\frac{24\eta d L^2}{\alpha}.
\end{align}
\end{restatable}
The proof is given in \cref{sec:proof-qsvrg-lmc}.

\begin{restatable}[Main Theorem for \algname{QSVRG-LMC}]{theorem}{qlmc}
\label{thm:qlmc}
    Let $\mu_k$ be the distribution of $\x_{k}$ in \algname{QSVRG-LMC} algorithm. Suppose that $f$ satisfies \cref{assumption:component-smoothness,assumption:LSI}.  Then for $\eta = \Ocal\left(\frac{\epsilon \alpha }{dL^2} \wedge \frac{\alpha}{L^2 m}\right)$, $K = \Tilde{\Ocal}\left(\frac{L^2\log(\KL(\mu_0||\pi)) }{\alpha^2}\left(n^{2/3} + \frac{d}{\epsilon}\right) \right)$, $b = \Tilde{\Ocal}(n^{1/3})$, and $m=\Tilde{\Ocal}(n^{2/3})$
    we have
    \[  
    \left\{\KL(\mu_{K}||\pi), \TV(\mu_{K},\pi)^2,\frac{\alpha}{2}\W_2(\mu_{K},\pi)^2\right\} \leq \epsilon.
    \]
The stochastic-gradient oracle query complexity is
$\Tilde{\Ocal}\!\left(
\frac{L^2\log(\KL(\mu_0\Vert\pi))}{\alpha^2}
\left(nd^{1/2}+\frac{d^{3/2}n^{1/3}}{\epsilon}\right)
\right)$.
\end{restatable}
\begin{proof}
    Setting $b = \Tilde{\Ocal}(n^{1/3})$ and $m = \Tilde{\Ocal}(n^{2/3})$ and $\eta \leq \frac{\epsilon \alpha }{176 dL^2}$
  the second term on the right hand side of   \cref{thm:qsvrgconvergence} becomes smaller than $\epsilon/2$. By the step size requirement of \cref{thm:qsvrgconvergence}, we have $\eta \leq \frac{\epsilon \alpha }{176 dL^2} \wedge \frac{\alpha}{24L^2 m}$. The first term in  \cref{thm:qsvrgconvergence} is smaller than $\epsilon/2$ when $K \leq \frac{ \log(2\KL(\mu_0\|\pi )/\epsilon)}{\alpha \eta}$. Hence $\TV$ distance is smaller than $\epsilon$. The results for $\W_2$ distance and $\TV$ distance hold due to Talagrand's inequality ~\cite{OTTO2000361} and Pinsker's inequality~\cite{Tsybakov2009} respectively.
  The total gradient complexity is $bK = \Tilde{\Ocal}\left(\frac{L^2\KL(\mu_0\|\pi )}{\alpha^2}\left(nd^{1/2} + \frac{d^{3/2}n^{1/3}}{\epsilon}\right) \right) $.
\end{proof}

The query complexity in~\cite{NEURIPS2022_78e839f9} is \(\Ocal(n+n^{1/2}\epsilon^{-1})\), whereas ours is \(\Ocal(n+n^{1/3}\epsilon^{-1})\). The coupled term \(n^{1/3}\epsilon^{-1}\) is possible because the stochastic-gradient cost \(Kb\) and the full-gradient cost \(nK/m\) are balanced. In classical SVRG-LMC, the optimal choice is \(m=b=n^{1/2}\), where \(b\) is the inner batch size; in our algorithm, we have \(m=n^{2/3}\) and \(b=n^{1/3}\). Thus, quantum gradient estimators allow the exact gradient to be computed less frequently, which is the main mechanism behind the speedup.

\subsection{Sampling under Strong Convexity via Hamiltonian Monte Carlo}
\label{sec:hmc-overview}
Next, we consider quantum speedups for Hamiltonian Monte Carlo (HMC) for strongly convex potentials. We replace the stochastic gradients in HMC with quantum-estimated stochastic gradients based on the SVRG and CV estimators used in~\cref{algo:qsvrg-lmc,algo:qsvrg-cv}.

HMC augments the position $\x$ with an auxiliary momentum $\p$ and introduces the Hamiltonian $H(\x,\p)=f(\x)+\|\p\|^2/2$. Its continuous dynamics satisfy
\begin{equation}
\label{eq:Hamiltonian-dynamics}
\frac{\mathrm{d}\x}{\mathrm{d}t}=\frac{\partial H}{\partial\p},
\qquad
\frac{\mathrm{d}\p}{\mathrm{d}t}=-\frac{\partial H}{\partial\x}.
\end{equation}
The leapfrog discretization alternates position and momentum updates and refreshes the momentum between proposals. Stochastic-gradient HMC~\cite{pmlr-v32-cheni14} replaces exact gradients in these updates with stochastic estimates, as summarized below.

\begin{algorithm}[ht]
\begin{algorithmic}[1]
\caption{$\algname{SG-HMC}$}
\label{algo:HMC}
\INPUT Stochastic gradient routine $\g(\cdot)$, initial point $\x_0$, step size $\eta$, number of leapfrog steps $S$, number of proposals $T$.
\OUTPUT Approximate sample from $\pi\propto e^{-f(\x)}$.
\FOR{$t=0,\ldots,T-1$}
\STATE Set $\q_0=\x_t$ and sample $\p_0\sim\mathcal{N}(0,I_d)$.
\FOR{$s=0,\ldots,S-1$}
\STATE Set $\q_{s+1}=\q_s+\eta\p_s-\frac{\eta^2}{2}\g(\q_s)$.
\STATE Set $\p_{s+1}=\p_s-\frac{\eta}{2}\g(\q_s)-\frac{\eta}{2}\g(\q_{s+1})$.
\ENDFOR
\STATE Set $\x_{t+1}=\q_S$.
\ENDFOR
\STATE \textbf{Return} $\x_T$.
\end{algorithmic}
\end{algorithm}

The stochastic gradients in~\cref{algo:HMC} may be formed using subsampling, SVRG, or control variates. These methods lower the per-step cost but introduce variance that can degrade sample quality and slow convergence; our quantum estimators reduce this variance at lower query cost.

\begin{assumption}[Strong Convexity]
There exists a positive constant $\mu$ such that for all $\x, \y \in \mathbb{R}^d$ it holds that
\label{assumption:strong-convexity}
    \begin{equation}
        f(\x)\geq  f(\y) + \langle \nabla f(\y), \x-\y \rangle +\frac{\mu}{2}\|\x-\y\|^2.
    \end{equation}
\end{assumption}
For smooth objectives, we occasionally use the condition number \(\kappa=L/\mu\) to express the dependence on \(L\) and \(\mu\) more compactly.

Based on \cref{assumption:strong-convexity,assumption:component-smoothness}, we give the main theorem for the quantum Hamiltonian Monte Carlo algorithm implemented with the quantum SVRG estimator.
\begin{restatable}[Main Theorem for \algname{QSVRG-HMC}]{theorem}{qsvrg}
\label{thm:qhmc-svrg-main}
    Let $\mu_k$ be the distribution of $\x_{k}$ in \algname{QSVRG-HMC} algorithm. Suppose that $f$ satisfies \cref{assumption:strong-convexity,assumption:component-smoothness}. Given that the initial point $\x_0$ satisfies $\|\x_0 - \arg\min_{\x} f(\x)\|\leq \frac{d}{\mu}$, then, for $\eta = \Ocal\left(\frac{\epsilon}{L^{1/2}d^{1/2} \kappa^{3/2}} \right)$, $S = \Tilde{\Ocal}\left(\frac{Ld^{1/2}\kappa^{3/2}}{\epsilon}\right)$, $T=\Tilde{\Ocal}(1)$,  $b = \Ocal \left(\frac{L^{1/8}\epsilon^{1/4}n^{1/2}}{d^{1/8}\kappa^{3/8}} \vee 1\right)$, and $m = n/b$, we have
    \[
    \W_2(\mu_{ST},\pi) \leq \epsilon.
    \]
    The total query complexity to the stochastic gradient oracle is $\Tilde{\mathcal{O}}\left(\frac{Ld^{1/2}\kappa^{3/2}}{\epsilon}+\frac{L^{9/8}d^{7/8}\kappa^{3/4}n^{1/2}}{\epsilon^{3/4}} \right)$. 
\end{restatable}
\begin{proof}
By the choice of $\eta$ in the theorem statement and the variance upper bound in \cref{lem:modified-variance-lemma}, $\eta = \Ocal(L^{1/2}\sigma^{-2}\kappa^{-1}\wedge L^{-1/2} )$. Therefore, by \cref{thm:HMC-convergence}, for $K = \frac{1}{4\sqrt{L}\eta}$, we have
\begin{equation}
\label{eq:svrg-w2-equation}
 \W_2(\mu_T, \pi)\leq (1-(128\kappa)^{-1})^{\frac{T}{2}}(2D+ 2d/\mu)^{1/2} + \Gamma_1\eta^{1/2}+\Gamma_2\eta
\end{equation}
where,
\begin{align}
    \Gamma_1^2 &= \Ocal\left(\frac{L^{1/2} m^2 \kappa^3 d \eta^2}{b^2}\right),\\
    \Gamma_2^2 &= \Ocal\left(\kappa^3d + \frac{L^{3/2} m^2 \kappa^3 d \eta^3}{b^2}\right).
\end{align}
We set $b m =  \Ocal(n)$. The first term in \cref{eq:svrg-w2-equation} is $\Ocal(\epsilon)$ when $T = \Tilde{\Ocal}(\log(1/\epsilon))$.
The last two terms in \cref{eq:HMC-convergence} for \algname{QHMC-SVRG} become $\Ocal\left(\frac{L^{1/4} d^{1/2}\kappa^{3/2}\eta^{3/2} n  }{b^2} + d^{1/2}\kappa^{3/2}  \eta\right)$. For $b = \Ocal(d^{-1/8}\kappa^{-3/8}\epsilon^{1/4}n^{1/2}L^{1/8} \vee 1)$ and $\eta = \Ocal(\epsilon\kappa^{-3/2}d^{-1/2} )$, the bias term becomes $\Ocal(\epsilon)$. Using \cref{lem:var-reduction}, the number of gradient calculations scales as $\Tilde{O}(Ld^{1/2}\kappa^{3/2}\epsilon^{-1}+L^{9/8}d^{7/8}\kappa^{3/4}\epsilon^{-3/4}n^{1/2} )$.
\end{proof}

We obtain an analogous guarantee for HMC with the quantum control-variate estimator.

\begin{restatable}[Main Theorem for \algname{QCV-HMC}]{theorem}{qcv}
\label{thm:qhmc-cv-main}
    Let $\mu_k$ be the distribution of $\x_{k}$ in \algname{QCV-HMC} algorithm. Suppose that $f$ satisfies \cref{assumption:strong-convexity,assumption:component-smoothness}. Given that the initial point $\x_0$ satisfies $\|\x_0 - \arg\min_{\x} f(\x)\|\leq \frac{d}{\mu}$, then, for $\eta = \Ocal\left(\frac{\epsilon}{L^{1/2}d^{1/2} \kappa^{3/2}} \right)$, $S = \Tilde{\Ocal}\left(\frac{Ld^{1/2}\kappa^{3/2}}{\epsilon}\right)$, $T=\Tilde{\Ocal}(1)$, and  $b = \Ocal\left(\frac{d^{1/4}\kappa^{3/4}}{L^{1/4}\epsilon^{1/2}} \vee 1\right)$, we have
    \[
    \W_2(\mu_{ST},\pi) \leq \epsilon.
    \]
   The total query complexity to the stochastic gradient oracle is $\Tilde{\mathcal{O}}\left(\frac{Ld^{5/4}\kappa^{9/4}}{\epsilon^{3/2}}\right)$.
\end{restatable}
\begin{proof}
By the choice of $\eta$ in the theorem statement and the variance upper bound in \cref{lem:modified-variance-lemma2}, $\eta = \Ocal(L^{1/2}\sigma^{-2}\kappa^{-1}\wedge L^{-1/2} )$. Therefore, by \cref{thm:HMC-convergence}, for $K = \frac{1}{4\sqrt{L}\eta}$, we have
\begin{equation}
\label{eq:cv-w2-equation}
 \W_2(\mu_T, \pi)\leq (1-(128\kappa)^{-1})^{\frac{T}{2}}(2D+ 2d/\mu)^{1/2} + \Gamma_1\eta^{1/2}+\Gamma_2\eta,
\end{equation}
where,
\begin{align}
  \Gamma_1 &= \Ocal\left(\frac{L^{-1/2}\kappa^3 d }{b^2}\right),\\
\Gamma_2 &=  \Ocal\left(\kappa^3 d\right). 
\end{align}
The first term in \cref{eq:cv-w2-equation} is $\Ocal(\epsilon)$ when $T = \Tilde{\Ocal}(1)$.
The last two terms in  \cref{eq:cv-w2-equation} for \algname{QHMC-CV} become $\Ocal\left(\frac{L^{-1/4} d^{1/2}\kappa^{3/2}\eta^{1/2}}{b^2} + d^{1/2}\kappa^{3/2}  \eta\right)$. For $b = \Ocal(L^{-1/4}d^{1/4}\kappa^{3/4}\epsilon^{-1/2} \vee 1)$ and $\eta = \Ocal(\epsilon d^{-1/2}\kappa^{-3/2} )$, the bias term becomes $\Ocal(\epsilon)$. Using \cref{lem:var-reduction}, the number of gradient calculations scales as $\Tilde{O}(Ld^{1/2}\kappa^{3/2}\epsilon^{-1}+L^{3/4}d^{5/4}\kappa^{9/4}\epsilon^{-3/2} ) = \Tilde{O}(Ld^{5/4}\kappa^{9/4}\epsilon^{-3/2} )$.
\end{proof}

The proofs follow the same idea as in the LMC setting. We express the stochastic-gradient variance \(\sigma\) along the HMC trajectory in terms of \(b\) and the distance between the current iterate and the last iterate at which the full gradient was computed. We then choose \(b\) so that the per-iteration cost of quantum mean estimation, \(\Tilde{\Ocal}(\sigma/\hat{\sigma})\), balances the amortized full-gradient calculation cost, \(\Tilde{\Ocal}(n/m)\), since the full gradient is computed only once every \(m\) iterations.
\begin{table*}[ht!]
  \caption{Summary of the results in the stochastic gradient oracle setting (some of the previous results use a different scaling of $f$ and we convert the results to the same scaling as ours in the table). Here, we mainly focus on $n$ and $\epsilon$ dependency. See \cref{thm:qhmc-svrg-main,thm:qhmc-cv-main,thm:qlmc} for explicit dependencies on $L,\mu,\alpha,d$.}
\begin{center}
    
\small
\begin{tabular}{ |c|c|c|c| } 
\hline
 Algorithm & Assumptions  & Metric& Gradient Complexity\\
\hline
SG-HMC \cite{pmlr-v139-zou21b}& Strongly Convex & $\W_2$ &$ \tilde{\Ocal}(n\epsilon^{-2})$    \\ 
SVRG-HMC \cite{pmlr-v139-zou21b}& Strongly Convex & $\W_2$ &$ \tilde{\Ocal}(n^{2/3}\epsilon^{-2/3}+\epsilon^{-1})$    \\ 
SAGA-HMC \cite{pmlr-v139-zou21b}& Strongly Convex & $\W_2$ &$ \tilde{\Ocal}(n^{2/3}\epsilon^{-2/3}+\epsilon^{-1})$    \\ 
CV-HMC \cite{pmlr-v139-zou21b}& Strongly Convex &  $\W_2$ &$ \tilde{\Ocal}(\epsilon^{-2})$    \\ 
SRVR-HMC \cite{NEURIPS2019_c3535feb}&Dissipative Gradients & $\W_2$& $\tilde{\Ocal}(n+n^{1/2}\epsilon^{-2}+\epsilon^{-4} )$ \\
SVRG-LMC \cite{NEURIPS2022_78e839f9}& LSI& $\KL$ &$\tilde{\Ocal}(n+n^{1/2}\epsilon^{-1})$ \\
SARAH-LMC \cite{NEURIPS2022_78e839f9}& LSI& $\KL$ &$\tilde{\Ocal}(n+n^{1/2}\epsilon^{-1})$ \\
\hline
QSVRG-HMC [\cref{thm:qhmc-svrg-main}]& Strongly Convex & $\W_2$ & $\Tilde{\Ocal}(n^{1/2}\epsilon^{-3/4}+\epsilon^{-1})$\\
QCV-HMC [\cref{thm:qhmc-cv-main}]& Strongly Convex & $\W_2$ & $\Tilde{\Ocal}(\epsilon^{-3/2})$\\
QSVRG-LMC [\cref{thm:qlmc}]& LSI & $\KL \footnotemark$ & $\tilde{\Ocal}(n+n^{1/3}\epsilon^{-1})$\\
\hline
\end{tabular}
\end{center}
\label{table:results}
\end{table*}
\footnotetext{Convergence in $\KL$ divergence implies convergence in squared $\TV$ and $\W_2$ distances due to Pinsker's and Talagrand's inequalities. }

\cref{thm:qhmc-svrg-main,thm:qhmc-cv-main} imply that when $n=\Ocal(\epsilon^{-1/2})$ the best classical ($\algname{SVRG-HMC}$) and the best quantum ($\algname{QSVRG-HMC}$) algorithms have  $\Tilde{\Ocal}(\epsilon^{-1})$ gradient complexity.
On the other hand, when \(n=\omega(\epsilon^{-1})\), quantum algorithms have better complexity than the best classical algorithms; the better of \(\algname{QSVRG-HMC}\) and \(\algname{QCV-HMC}\) depends on the size of \(n\).

\begin{remark}
    Both the classical algorithms in \cite{pmlr-v139-zou21b} and quantum algorithms in this paper assume that the starting point is ($d/\mu$)-close to the minimizer $\x^{\star}=\arg \min f(\x)$. In case this point is not given, it can be obtained using $\Ocal(n)$ iterations of SGD \cite{10.1007/s11222-018-9826-2}.
\end{remark}

\section{Quantum Gradient Estimation in Zeroth-Order Stochastic Setting}
\label{sec:zeroth-order-gradient}
We characterize the complexity of our algorithms in this section with respect to the stochastic zeroth-order oracle defined in~\cref{eq:binary-oracle}. The construction uses Jordan's quantum gradient estimator and MLMC debiasing; these algorithms are reviewed in~\cref{sec:Jordan,sec:mlmc-overview}, with the full procedures given in~\cref{algo:grad_est,algo:MLMC}.
\subsection{Jordan's Gradient Estimation Algorithm}
\label{sec:Jordan}
Jordan's algorithm~\cite{Jordan_2005} approximates a gradient on a small grid around the point of interest and encodes the estimate into a quantum phase. It then applies an inverse quantum Fourier transform to recover all gradient coordinates. Gily\'en, Arunachalam, and Wiebe~\cite{Gily_n_2019} extended the original analysis to functions in the Gevrey class using central-difference formulas and a binary oracle model. The analysis closest to our setting is due to~\cite{Chakrabarti2020quantumalgorithms}, which gives constant query complexity for functions with Lipschitz gradients when function values can be queried with sufficient precision.

The following lemma from~\cite{Chakrabarti2020quantumalgorithms} shows that~\cref{algo:grad_est} has $\Tilde{O}(1)$ query complexity for a smooth function with high probability.

\begin{lemma}[Lemma 2.2 in~\cite{Chakrabarti2020quantumalgorithms}]
\label{lem:jordan_whp}
Let $f\colon\mathbb{R}^d\to\mathbb{R}$ be accessible through an evaluation oracle with error at most $\epsilon$. Assume that $\|\nabla f\|\leq L$ and that $f$ is $\beta$-smooth in $B_\infty(x,2\sqrt{\epsilon/\beta})$. Let $\tilde{\g}$ be the output of $\algname{QuantumGradient}(f,\epsilon,M,\beta,\x_0)$ from~\cref{algo:grad_est}. Then
\begin{align}
\Pr\left[|\tilde{\g}_i-\nabla f(\x)_i|>1500\sqrt{d\epsilon\beta}\right]<\frac13,
\qquad \forall i\in[d].
\end{align}
\end{lemma}

Repeating~\cref{algo:grad_est} and taking the coordinate-wise median gives a smooth high-probability estimate when the gradient norm is bounded~\cite[Lemma~2.3]{Chakrabarti2020quantumalgorithms}. Achieving $L_2$ error $\delta$ requires evaluation error $\Ocal(\delta^2/d^2)$, which is difficult when the evaluation oracle is stochastic. Our robust phase oracle in~\cref{prop:phase-oracle} addresses this issue for noisy function values.

\begin{algorithm}[H]
\begin{algorithmic}[1]
  \caption{$\algname{QuantumGradient}(f,\epsilon,L,\beta,x_0)$}
  \label{algo:grad_est}
  \item \textbf{Input}: {Function $f$, evaluation error $\epsilon$, gradient norm bound $L$, smoothness parameter $\beta$, and point $\x_0$.\\
    Define
    \begin{itemize}
      \item $l = 2\sqrt{{\epsilon}/{\beta d}}$ to be the size of the grid used,
      \item $b \in \mathbb{N}$ such that $\frac{24\pi\sqrt{d\epsilon \beta}}{L} \le \frac{1}{2^b} = \frac{1}{N} \le \frac{48\pi\sqrt{d\epsilon \beta}}{L}$,
      \item $b_0 \in \mathbb{N}$ such that $\frac{N\epsilon}{2Ll} \le \frac{1}{2^{b_0}} = \frac{1}{N_0} \le \frac{N\epsilon}{Ll}$,
      \item $F(x) = \frac{N}{2Ll}[f(x_0 + \frac{l}{N}(x - N/2)) - f(x_0)]$, and
      \item $ \gamma: \{0,1,\dots,N-1\} \to G:=\{-N/2,-N/2+1,\dots,N/2-1\}$ such that $\gamma(x) = x - N/2$.
    \end{itemize}
    Let $O_F$ denote a unitary operation acting as $O_F\ket{x} = e^{2\pi i \tilde{F}(x)}\ket{x}$, where $|\tilde{F}(x) - F(x) | \le \frac{1}{N_0}$, with $x$ represented using $b$ bits and $\tilde{F}(x)$ represented using $b_0$ bits.
  }
  \STATE Start with $d$ $b$-bit registers set to zero and apply a Hadamard transform to each to obtain
  \begin{align*}
    \frac{1}{\sqrt{N^d}}\sum_{x_1,\ldots,x_d \in \{0,1,\ldots,N-1\}}\ket{x_1,\ldots,x_d}.
  \end{align*}
  \STATE Apply $O_F$ and the map $\ket{x} \mapsto \ket{\gamma(x)}$ to obtain
  \begin{align*}
    \frac{1}{N^{d/2}}\sum_{\g \in G^d}e^{2\pi i \tilde{F}(\g)} \ket{\g}.
  \end{align*}
  \STATE Apply the inverse QFT over $G$ to each register.
  \STATE Measure the final state to get $k_1,k_2,\ldots,k_d$ and report $\tilde{\g} = \frac{2L}{N}(k_1,k_2,\ldots,k_d)$. \label{lin:grad_est_measure}
\end{algorithmic}
\end{algorithm}

\subsection{Multilevel Monte Carlo Algorithm}
\label{sec:mlmc-overview}
Jordan's estimator is biased. We use multilevel Monte Carlo (MLMC) to turn an increasingly accurate family of biased stochastic-gradient estimators into an unbiased estimator without changing the leading query complexity. Suppose $\algname{BiasedStochasticGradient}(\x,\hat{\sigma})$ has mean-square error at most $\hat{\sigma}^2$ and cost $\Tilde{\Ocal}(C/\hat{\sigma})$. The following randomized telescoping construction performs the debiasing.

\begin{algorithm}[H]
\begin{algorithmic}[1]
  \caption{$\algname{UnbiasedStochasticGradient}$}
  \label{algo:MLMC}
  \INPUT Estimator $\algname{BiasedStochasticGradient}$ and target variance $\hat{\sigma}^2$.
  \OUTPUT Unbiased estimate $\g$ of $\nabla f(\x)$ with variance at most $\hat{\sigma}^2$.
  \STATE Set $\g_0\leftarrow\algname{BiasedStochasticGradient}(\x,\hat{\sigma}/10)$.
  \STATE Sample $j\sim\mathrm{Geom}(1/2)\in\mathbb{N}$.
  \STATE Set $\g_j\leftarrow\algname{BiasedStochasticGradient}(\x,2^{-3j/4}\hat{\sigma}/10)$.
  \STATE Set $\g_{j-1}\leftarrow\algname{BiasedStochasticGradient}(\x,2^{-3(j-1)/4}\hat{\sigma}/10)$.
  \STATE Set $\g\leftarrow\g_0+2^j(\g_j-\g_{j-1})$.
  \STATE \textbf{Return} $\g$.
\end{algorithmic}
\end{algorithm}

\begin{lemma}
\label{lem:mlmc}
Given access to an algorithm $\algname{BiasedStochasticGradient}$ that outputs a random vector $\mathbf{v}$ with $\mathbb{E}\|\mathbf{v}-\nabla f(\x)\|^2\leq\hat{\sigma}^2$ at cost $\Tilde{\Ocal}(C/\hat{\sigma})$, \cref{algo:MLMC} outputs a vector $\g$ such that $\mathbb{E}[\g]=\nabla f(\x)$ and $\mathbb{E}\|\g-\nabla f(\x)\|^2\leq\hat{\sigma}^2$ with expected cost $\Tilde{\Ocal}(C/\hat{\sigma})$.
\end{lemma}

\begin{proof}
We adopt the proof in \cite{NEURIPS2023_6ed9931d}. We start by writing

\begin{align}
\g=\g_0+2^J(\g_J-\g_{J-1}),\qquad J\sim\mathrm{Geom}\Big(\frac{1}{2}\Big)\in\mathbb{N}.
\end{align}
Given that $\Pr(J=j)=2^{-j}$, we have
\begin{align}
\mathbb{E}[\g]=\mathbb{E}[\g_0]+\sum_{j=1}^{\infty}\Pr(J=j)2^j(\mathbb{E}[\g_j]-\mathbb{E}[\g_{j-1}])
=\mathbb{E}[\g_{\infty}]=\nabla f(\x).
\end{align}
As for the variance, using the inequality $(a+b)^2\leq 2a^2+2b^2$, we have
\begin{align}
\mathbb{E}\|\g-\nabla f(\x)\|^2&\leq 2\mathbb{E}\|\g-\g_0\|^2+2\mathbb{E}\|\g_0-\nabla f(\x)\|^2
\end{align}
where
\begin{align}
\mathbb{E}\|\g-\g_0\|^2
=\sum_{j=1}^{\infty}\Pr(J=j)2^{2j}\mathbb{E}\|\g_j-\g_{j-1}\|^2
=\sum_{j=1}^{\infty}2^j\mathbb{E}\|\g_j-\g_{j-1}\|^2,
\end{align}
and for each $j$ we have 
\begin{align}
\mathbb{E}\|\g_j-\g_{j-1}\|^2
\leq 2\mathbb{E}\|\g_j-\nabla f(\x)\|^2+2\mathbb{E}\|\g_{j-1}-\nabla f(\x)\|^2.
\end{align}
By assumption on $\algname{BiasedStochasticGradient}$,
\begin{align}
\mathbb{E}\|\g_j-\nabla f(\x)\|^2\leq\frac{\hat{\sigma}^2}{100\cdot 2^{3j/2}},\quad\forall j\geq 0,
\end{align}
which leads to 
\begin{align}
\mathbb{E}\|\g_j-\g_{j-1}\|^2\leq \frac{\hat{\sigma}^2}{50\cdot 2^{3(j-1)/2}}+\frac{\hat{\sigma}^2}{50\cdot 2^{3j/2}}\leq\frac{\hat{\sigma}^2}{10\cdot 2^{3j/2}},
\end{align}
and
\begin{align}
\mathbb{E}\|\g-\g_0\|^2=\frac{\hat{\sigma}^2}{10}\sum_{j=1}^{\infty}\frac{1}{2^{j/2}}\leq \frac{1}{3} \hat{\sigma}^2\,.
\end{align}
Hence,
\begin{align}
\mathbb{E}\|\g-\nabla f(\x)\|^2\leq 2\mathbb{E}\|\g-\g_0\|^2+2\mathbb{E}\|\g_0-\nabla f(\x)\|^2\leq \hat{\sigma}^2,
\end{align}
Moreover, the expected cost is
\begin{align}
\Tilde{\Ocal}\left(\frac{C}{\hat{\sigma}}\right)\cdot\left(1+\sum_{j=1}^{\infty}\Pr\{J=j\}\cdot\left(2^{3j/4}+2^{3(j-1)/4}\right)\right)=\Tilde{\Ocal}\left(\frac{C}{\hat{\sigma}}\right).
\end{align}
\end{proof}

\subsection{Gradient Estimation for Smooth Potentials}
We assume the following about the evaluation oracle for $f$.
\begin{assumption}[Bounded Noise]
\label{ass:bounded-variance-zeroth-order}
For any \(\mathbf{x} \in \mathbb{R}^d\), the stochastic zeroth-order oracle outputs an estimator \( f(\mathbf{x}; \xi) \) of \( f(\mathbf{x}) \) such that $\mathbb{E} [f (\mathbf{x}; \xi)] = f (\mathbf{x})$, $\mathbb{E} [\nabla f (\mathbf{x}; \xi)] = \nabla f (\mathbf{x})$, and $\mathbb{E} \|\nabla f (\mathbf{x}; \xi) - \nabla f (\mathbf{x})\|^2 \leq \sigma^2$.
\end{assumption}

We note that gradients of \(f(\cdot;\xi)\) are not queried by the algorithm and only appear in the assumption.

\begin{assumption}[Smoothness]
\label{ass:full-smoothness}
The potential function \(f: \mathbb{R}^d \to \mathbb{R}\) has \(L\)-Lipschitz gradients. Specifically, for any $\x,\y \in \mathbb{R}^d$ it holds that
\[
\|\nabla f(\x) - \nabla f(\y)\| \leq L \|\x - \y\|. 
\]
\end{assumption}

The assumption is broader than an additive-noise model because it also covers certain multiplicative-noise models. For example, suppose \(f:\mathcal{B}_d(0,R)\to\mathbb{R}\) is \(L\)-smooth and differentiable, and the stochastic components have the form \(f(\x;\xi)=\xi f(\x)\), where \(\mathbb{E}[\xi]=1\) and $\mathbb{E}[(\xi-1)^2] \leq \frac{\sigma^2}{4L^2R^2}$. Then \cref{ass:bounded-variance-zeroth-order} is satisfied.

We assume that \(f\) can be queried with the same randomness at two points, so that \(f(\x;\xi_i)\) and \(f(\y;\xi_i)\) can be evaluated using the same seed $\xi_i$. In the finite-sum case, for instance, one can prepare \(\frac1{\sqrt n}\sum_{i=1}^n\ket{0}\ket{i}\) and then, using the binary oracle, prepare \(\frac1{\sqrt n}\sum_{i=1}^n\ket{f(\x;i)}\ket{i}\). This requires no QRAM-style data encoding, since the superposition is only over the indices. 
Classically, the gradient in this two-point setting can be estimated using the Gaussian smoothing technique. This involves sampling random directions from the extended space around the target point and performing two-point evaluations to approximate the gradient. Specifically, the gradient can be approximated as:
\begin{align}\label{eq:gradest}
    \g_{\nu,b}(\x) = \frac{1}{b} \sum_{i=1}^b \frac{f(\x + \nu \uu_i; \xi_i) - f(\x; \xi_i)}{\nu} \uu_i,
\end{align}
where \( \uu_i \sim \mathcal{N}(0, I_d) \) are independent and identically distributed random vectors. Authors in~\cite{10.1007/s10208-021-09499-8} showed that for any $\x \in \mathbb{R}^d$, the estimator $\g_{\nu,b}$ satisfies $\mathbb{E}\|\g_{\nu,b}(\x) - \nabla f(\x)\|^2 
        \leq \frac{4(d+5)(\|\nabla f(\x)\|^2 + \sigma^2)}{b} + \frac{3\nu^2 L^2 (d+3)^3}{2}$. Although the squared norm of the gradient on the right-hand side is unbounded, it is typically of order \(\Tilde{\mathcal{O}}(d)\) in expectation throughout the trajectory of LMC. Consequently, this method requires \( b = \mathcal{O}(d^2 / \epsilon^2) \) function evaluations to achieve an \(\epsilon\)-accurate gradient estimate in the \( L_2 \) norm.

To our knowledge, no existing quantum algorithm estimates the gradient in this particular setting. Our goal is to develop additional tools that allow Jordan's quantum estimation algorithm~\cite{Jordan_2005} to be used under this oracle model. The main technical tool is the following robust phase oracle.

\begin{restatable}{proposition}{phaseoracle}
    \label{prop:phase-oracle}
    Let $X \in \mathbb{R}$ be a random variable such that $\mathbb{E}\|X - \mathbb{E}[X]\|^2\leq \sigma^2 $. Given two reals $t\geq 0$ and $\epsilon \in (0,1)$, then there is a unitary operator $P^{X}_{t,\epsilon}: \ket{0}\ket{0}\mapsto \ket
    {\phi_X}\ket{0}$ acting on $\mathcal{H}_X\otimes \mathcal{H}_{\text{aux}}$ that can be implemented using $\Tilde{\Ocal}(t \sigma \log(1/\epsilon))$ quantum experiments and binary oracle queries to $X$ such that 
    \[
    \|\ket{\phi_X} - e^{it \mathbb{E}[X]}\ket{0}\|\leq \epsilon,
    \]
    with probability at least $8/9$.
\end{restatable}
This phase oracle is similar to the oracle implemented in~\cite{Cornelissen_2022}, but their construction requires \(\|X\|\leq 1\), whereas \(X\) may be unbounded in our setting. Thus \cref{prop:phase-oracle} extends the phase oracle to unbounded random variables by using a sequence of unitaries at different truncation levels. In our gradient estimation algorithm, this oracle replaces the phase oracle in~\cref{algo:grad_est}, with \(X=f(\x;\xi)\). Since Jordan's algorithm is biased and succeeds with high probability, we postprocess its output using a multilevel Monte Carlo algorithm (See ~\cref{sec:mlmc-overview} for overview) to obtain a smooth unbiased gradient estimator.
The proof of \cref{prop:phase-oracle} is given in \cref{sec:gradient-estimation-appendix}.

\begin{restatable}{theorem}{smoothgradientwithphase}
    \label{thm:zeroth-order-gradient1}
    Suppose that the potential function $f$ satisfies \cref{ass:full-smoothness,ass:bounded-variance-zeroth-order} and further suppose that $\|\nabla f(\x)\|\leq M$\footnote{One can show that the norm of the gradient is bounded by a function of problem parameters in such a large ball throughout the trajectory of HMC or LMC due to smoothness. Since the dependency on $M$ is logarithmic, we do not give an explicit bound on $M$.} for all $\x \in \mathcal{B}(x^{\star}, \textup{poly}(d))$. Then, given a real $\x \in \mathcal{B}(x^{\star}, \textup{poly}(d))$ and $\hat{\sigma}>0$, there exists a quantum algorithm that outputs a random vector $\g$ such that 
    \[
        \mathbb{E}[\g]= \nabla f(\x), \quad \text{and} \quad \mathbb{E}\|\g- \nabla f(\x)\|^2\leq \hat{\sigma}^2
    \]
    using $\Tilde{O}(\frac{\sigma d}{\hat{\sigma}})$ queries to the stochastic evaluation oracle.
\end{restatable}
\begin{proof}
    Suppose that we run \cref{algo:grad_est} in \cref{lem:zeroth-order-gradient1} $T$ times with target accuracy $\frac{\hat{\sigma}}{2}$, then compute the median (coordinate-wise) of these outputs. If the result has norm smaller than $M$, we output this vector. Otherwise, we output the all-$0$ vector. Let $\mathbf{v}$ be the output of this algorithm. Since the algorithm in \cref{lem:zeroth-order-gradient1} outputs a vector $\Tilde{\g}$ such that $\|\Tilde{\g} - \nabla f(\x)\|\leq \frac{\hat{\sigma}}{2}$ with high probability, then by Chernoff bound and union bound over each dimension, at least $\frac{T}{2}$ of the outputs satisfy $\|\Tilde{\g} - \nabla f(\x)\|\leq \hat{\sigma}$ with probability at least $1-2\exp(-T^2/24)$. Since the norm of the gradient is $M$, when the condition fails, the error is $\|\Tilde{\g}-\nabla f(\x)\|\leq M$.  Then in expectation,
    \begin{equation}
        \mathbb{E}\|\mathbf{v} - \nabla f(\x)\|^2 \leq \frac{\hat{\sigma}^2}{4} + 2\exp(-T^2/24)M^2.
    \end{equation}
    Setting $T^{2} = 24 \log (\frac{8M^2}{3\hat{\sigma}^2 })$ gives $  \mathbb{E}\|\mathbf{v} - \nabla f(\x)\|^2 \leq \hat{\sigma}^2$. Hence, the overhead to \cref{lem:zeroth-order-gradient1} to make the output smooth is at most logarithmic. Finally, we can use this algorithm as the biased stochastic gradient estimator in \cref{algo:MLMC} and obtain an unbiased estimator $\g$.
\end{proof}

Although~\cref{thm:zeroth-order-gradient1} is motivated by the sampling task, it might be of independent interest for various optimization problems even in non-smooth settings. For example, suppose that $f_{\xi}$ is a non-smooth but locally $L$-Lipschitz function on the grid $G_d^l$. We define $f_v(\x) =\mathbb{E}_{\xi\in \Xi, \uu\sim \mathcal{B}(0,1)}[f(\x+v\uu;\xi)]$. Then, let $\y \in G_d^l$, and we have $\mathbb{E}\|\nabla f(\y+v\uu) - \nabla f_v(\y) \|^2\leq 4L^2$. It is known that $f_v$ is a smooth function with smoothness parameter $O(Ld^{1/2}v^{-1})$. Hence, by \cref{thm:zeroth-order-gradient1} our algorithm outputs an unbiased estimator $\g$ such that $\mathbb{E}[\g] = \nabla f_{v}(\x)$ and $
\mathbb{E}\|\g - \nabla f_v(\x)\|^2\leq \hat{\sigma}^2$ using $\Tilde{\Ocal}(\frac{L d}{\hat{\sigma}})$ queries to $f_{\xi}$. This result has recently been established in \cite{liu2024quantum}, and it is a special case of \cref{thm:zeroth-order-gradient1}. Hence, our quantum gradient estimation can give speedups beyond the settings considered in \cite{liu2024quantum}, but this is outside the scope of this work.

\subsection{Gradient Estimation under Additional Smoothness Assumption}
\label{sec:grad-est-details}
In this section, we impose a slightly stronger smoothness assumption on the stochastic functions \(f_\xi\), which further improves the dimension dependence.

\begin{assumption}[Lipschitz Stochastic Gradients]
\label{ass:smoothness}
The stochastic component \(f(\cdot; \xi): \mathbb{R}^d \to \mathbb{R}\) has \(L(\xi)\)-Lipschitz gradients for any \(\xi \in \Xi\). Specifically, it holds that 
\begin{align}
    \|\nabla f(\x; \xi) - \nabla f(\y; \xi)\| \leq L(\xi) \|\x - \y\|, 
\end{align}
and the expected Lipschitz constant satisfies \(\mathbb{E}[L(\xi)] = L\).
\end{assumption}

\Cref{ass:smoothness} is weaker than assuming that each stochastic function \(f_\xi\) has \(L\)-Lipschitz gradients, and it is straightforward to show that \cref{ass:smoothness} implies that \(f\) has Lipschitz gradients.

In this section, we assume the following sampling oracle,
\begin{align}
    O_{\xi}:\ket{\x}\ket{0}\mapsto \sum_{\xi \in \Xi}\sqrt{\Pr(\xi)}\ket{\x}\ket{\xi}.
\end{align}
Instead of implementing an accurate phase oracle, one can first estimate \(\nabla f(\x;\xi)\) and then use quantum mean estimation to compute \(\nabla f(\x)\). Quantum mean estimation computes the mean of a random variable with quadratically fewer samples than classical mean estimation (\cref{lem:var-reduction}). 

However, \cref{ass:smoothness} implies that $f_{\xi}$ might not be a smooth function (even if $f$ is smooth), which is the requirement in~\cite{Chakrabarti2020quantumalgorithms}. Hence, Jordan's algorithm might fail to compute the gradient for $\nabla f_{\xi}$ with small probability regardless of the parameter choice in quantum gradient estimation. To address this, we propose a robust version of the quantum gradient estimation algorithm. Our final algorithm achieves \(\Tilde{\mathcal{O}}(d^{1/2} \epsilon^{-1})\) query complexity to estimate the gradient up to $\epsilon$ error.

\begin{algorithm}[htbp]
\begin{algorithmic}[1]
  \caption{$\algname{QuantumStochasticGradient}$}
  \label{algo:QuantumStochasticGradient}
  \item \textbf{Input}: {Access to stochastic functions $f_{\xi\in\Xi}$, variance $\sigma^2$, target $\epsilon$, smoothness parameter $L$, point $\x$.}\\
   Define $\beta = \frac{164 L\sigma^2}{\epsilon^2}$,
       $D = \frac{40 \sigma^2}{\epsilon}$,
       $\epsilon' = \frac{\epsilon^2}{\beta^2 d^3 (12000)^2}$.
\STATE Sample $\xi_0$ at random from $\Xi$.
\STATE Compute  $\boldsymbol{s} = \algname{QuantumGradient}(f_{\xi_0},\epsilon', M,\beta,\x)$ (See \cref{algo:grad_est}).

\STATE  Let $\mathcal{A}$ be a algorithm that runs $\g = \algname{QuantumGradient}(f_{\xi},\epsilon', M,\beta,\x)$ with random $\xi \in \Xi$ and outputs $\g$ if $\|\g-\boldsymbol{s}\|\leq D$, otherwise it outputs $\boldsymbol{s}$. Further suppose that $\mathcal{A}$ does not make any measurement.
\STATE Run $\mathcal{A}$ on the superposition state to obtain
$\sum_{\xi \in \Xi}\sqrt{\Pr(\xi)}\ket{\mathcal{A(\x)}}\ket{\xi}.$
\STATE Output $\boldsymbol{v} = \algname{QuantumMeanEstimation}(\mathcal{A}, \epsilon/4, \delta)$.
\end{algorithmic}
\end{algorithm}

The algorithm first samples \(\xi\), runs the gradient estimation algorithm, and stores the output \(\mathbf{s}\). The key idea is to use \(\mathbf{s}\) to replace outlier estimates and then show that this replacement does not significantly increase the error with high probability. Next, we run \(\mathcal{A}\) in superposition on \(\ket{\psi_1}=\sum_{\xi\in\Xi}\sqrt{\Pr(\xi)}\ket{\x}\ket{\xi}\). The final step of Jordan's algorithm produces the following quantum state:
\begin{align}
    \sum_{\xi \in \Xi}\sqrt{\Pr(\xi)}\ket{\Tilde{\g}(\x;\xi)}\ket{\x}\ket{\xi} + \ket{\mathcal{X}_1},
\end{align}
where \(\ket{\mathcal{X}_1}\) is another garbage state with a small amplitude and 
\begin{align}
    \Tilde{\g}(\x, \xi) = 
    \begin{cases} 
      \g(\x, \xi) & \text{if } \|\g(\x, \xi) - \mathbf{s}\| \leq D, \\
       \mathbf{s} & \text{otherwise}.
    \end{cases}
\end{align}
is the corrected gradient estimate.

Finally, we estimate the mean of the first register and output the resulting vector \(\boldsymbol{v}\) as the gradient estimate. The replacement step works because \(\mathbf{s}\) is within a controlled distance of the true gradient with high probability, so truncating large-error estimates changes the expectation by only a small amount. This truncation is nevertheless necessary because rare large errors can otherwise dominate the mean. A rigorous probabilistic analysis gives the following guarantee.

\begin{restatable}{lemma}{highprobgradient}
    \label{thm:high-prob-gradient}
    Under \cref{ass:smoothness,ass:bounded-variance-zeroth-order}, \cref{algo:QuantumStochasticGradient} returns a vector $\boldsymbol{v}$ such that 
    \begin{align}
        \|\boldsymbol{v}- \nabla f(\x)\|\leq \epsilon
    \end{align}
    with high probability using $\Tilde{\Ocal}(\sigma d^{1/2}\epsilon^{-1})$ queries to the stochastic evaluation oracle.
\end{restatable}
Next, we can postprocess the output of \cref{algo:QuantumStochasticGradient} using MLMC to obtain a smooth and unbiased estimate.
The proof of \cref{thm:high-prob-gradient} is given in \cref{sec:gradient-estimation-appendix}.

\begin{restatable}[Smooth Gradient]{theorem}{smoothgradient}
    \label{thm:smooth-gradient}
     Suppose that the potential function $f$ satisfies \cref{ass:smoothness,ass:bounded-variance-zeroth-order} and further suppose that $\|\nabla f(\x)\|\leq M$ for all $\x \in \mathcal{B}(x^{\star}, \textup{poly}(d))$. Then, given a real $\x \in \mathcal{B}(x^{\star}, \textup{poly}(d))$ and $\hat{\sigma}>0$, there exists a quantum algorithm that outputs a random vector $\g$ such that 
    \begin{equation}
         \mathbb{E}[\g]= \nabla f(\x), \quad \text{and} \quad \mathbb{E}\|\g- \nabla f(\x)\|^2\leq \hat{\sigma}^2   
    \end{equation}
    using $\Tilde{\Ocal}(\frac{\sigma d^{1/2}}{\hat{\sigma}})$ queries to the stochastic evaluation oracle in expectation.
\end{restatable}
\begin{proof}
    Suppose that we run \cref{algo:QuantumStochasticGradient} $T$ times with target accuracy $\frac{\hat{\sigma}}{2}$, then compute the median (coordinate-wise) of these outputs. If the result has norm smaller than $M$, we output this vector. Otherwise, we output all $0$ vector. Let $\mathbf{v}$ be the output of this algorithm. Since \cref{algo:QuantumStochasticGradient} outputs a gradient $\mathbf{v}$ such that $\|\mathbf{v} - \nabla f(\x)\|\leq \hat{\sigma}/2$ with high probability (say $2/3$), then by Chernoff bound and union bound over each dimension, at least $\frac{T}{2}$ of the outputs satisfy $\|\mathbf{v} - \nabla f(\x)\|\leq \hat{\sigma}$ with probability at least $1-2\exp(-T^2/24)$. Since the norm of the gradient is $M$, when the condition fails, the error is $\|\mathbf{v}-\nabla f(\x)\|\leq M$.  Then in expectation,
    \begin{equation}
        \mathbb{E}\|\mathbf{v} - \nabla f(\x)\|^2 \leq \frac{\hat{\sigma}^2}{4} + 2\exp(-T^2/24)M^2.
    \end{equation}
    Setting $T^{2} = 24 \log (\frac{8M^2}{3\hat{\sigma}^2 })$ gives $  \mathbb{E}\|\mathbf{v} - \nabla f(\x)\|^2 \leq \hat{\sigma}^2$. Hence, the overhead is at most logarithmic. Finally, we run \cref{algo:MLMC} to obtain an unbiased estimator $\g$.
\end{proof}

We remark that our focus is on query complexity. Quantum mean estimation and Jordan's algorithm use \(\Ocal(\operatorname{poly}(d,\log(1/\epsilon)))\) gates~\cite{Cornelissen_2022,chakrabarti2025speedupsconvexoptimizationquantum}, and our algorithm does not change this circuit structure.

\section{Quantum Speedups for Sampling via Evaluation Oracle}
\label{sec:sampling-via-zeroth-order}

We apply the quantum gradient estimation algorithms from the previous sections to establish convergence of HMC and LMC with inexact gradients in the strongly convex and LSI settings. The gradient estimation error creates a tradeoff: larger error increases the number of sampler iterations, while smaller error requires more evaluation queries. We optimize this error to minimize the total query cost. We first treat approximate sampling from strongly log-concave distributions using HMC, with error measured in the Wasserstein metric.

\begin{restatable}[Main Theorem for \algname{QZ-HMC}]{theorem}{qzhmc}
\label{thm:qzhmc}
    Let $\mu_k$ be the distribution of $\x_{k}$ in \algname{QZ-HMC} algorithm. Suppose that the~\cref{assumption:strong-convexity,ass:bounded-variance-zeroth-order,ass:full-smoothness} hold. Given that the initial point $\x_0$ satisfies $\|\x_0-\arg \min_{\x}f(\x)\|\leq \frac{d}{\mu}$, if we set the step size $\eta = \Ocal\left(\frac{\epsilon}{d^{1/2}\kappa^{3/2}}\right)$, $S = \Tilde{\Ocal}\left(\frac{Ld^{1/2}\kappa^{3/2}}{\epsilon}\right)$, $T=\Tilde{\Ocal}(1)$, and $\hat{\sigma}^2 = \Ocal\left(\frac{L^{3/2}d^{1/2}\epsilon}{\kappa^{3/2}}\right)$,  we have
    \[
    \W_2(\mu_{ST},\pi) \leq \epsilon.
    \]
     In addition, under \cref{ass:full-smoothness,ass:bounded-variance-zeroth-order}, the query complexity to the stochastic evaluation oracle is $\tilde{\Ocal}\left(\frac{d^{5/4}\sigma}{\epsilon^{3/2}}\right)$ or under \cref{ass:smoothness,ass:bounded-variance-zeroth-order} the query complexity to the stochastic evaluation oracle is $\tilde{\Ocal}\left(\frac{d^{3/4}\sigma}{\epsilon^{3/2}}\right)$.
\end{restatable}
\begin{proof}
By \cref{thm:HMC-convergence} for $\eta = \Ocal(L^{1/2}\sigma^{-2}\kappa^{-1}\wedge L^{-1/2} )$ and  $K = \frac{1}{4\sqrt{L}\eta}$, we have
\begin{equation}
 \W_2(\mu_T, \pi)\leq (1-(128\kappa)^{-1})^{\frac{T}{2}}(2D+ 2d/\mu)^{1/2} + \Gamma_1\eta^{1/2}+\Gamma_2\eta,
\end{equation}
where
\begin{align}
  \Gamma_1 &= \Ocal\left(L^{-3/2}\hat{\sigma}^2\kappa^2 \right),\\
\Gamma_2 &=  \Ocal\left(\kappa^3 d\right). 
\end{align}
The first term in error is $\Ocal(\epsilon)$ when $T = \Tilde{\Ocal}(\log(1/\epsilon))$.
The last two terms become $\Ocal\left(L^{-3/4} \hat{\sigma}\eta^{1/2} + d^{1/2}\kappa^{3/2}  \eta\right)$. For $\hat{\sigma} = \Ocal(L^{3/4}d^{1/4} \kappa^{-3/4}\epsilon^{1/2} \wedge \sigma)$ and $\eta = \Ocal(\epsilon d^{-1/2}\kappa^{-3/2} )$, the bias term becomes $\Ocal(\epsilon)$. Then, under \cref{ass:full-smoothness,ass:bounded-variance-zeroth-order}, the number of calls to evaluation oracle scale as $\Tilde{O}(d^{1/2}\kappa^{3/2}\epsilon^{-1}+\sigma d^{3/4}\kappa^{3/4}\epsilon^{-3/2} ) = \Tilde{\Ocal}(d^{3/4}\kappa^{3/4}\epsilon^{-3/2} )$. Similarly,  under \cref{ass:smoothness,ass:bounded-variance-zeroth-order} the evaluation complexity is $\Tilde{\Ocal}(\sigma d^{5/4}\kappa^{3/4}\epsilon^{-3/2} )$.
\end{proof}

Because strong convexity and the Wasserstein metric can be restrictive, we next consider non-convex potentials.

\begin{restatable}[Main Theorem for \algname{QZ-LMC}]{theorem}{qzlmc}
\label{thm:qzlmc}
    Suppose that the~\cref{assumption:strong-convexity,ass:bounded-variance-zeroth-order,ass:full-smoothness} hold. Let $\mu_k$ be the distribution of $\x_{k}$ in \algname{QZ-LMC} algorithm. Then, if we set the step size $\eta = \Ocal\left(\frac{\epsilon \alpha}{dL^2}\right)$, $K = \Tilde{\Ocal}\left(\frac{dL^2\log(\KL(\mu_0||\pi))}{\epsilon\alpha^2}\right)$, and $\hat{\sigma}^2 = \Ocal\left(\alpha \epsilon\right)$,
    we have
     \[  
    \left\{\KL(\mu_{K}||\pi), \TV(\mu_{K},\pi)^2,\frac{\alpha}{2}\W_2(\mu_{K},\pi)^2\right\} \leq \epsilon.
    \]
    In addition, under \cref{ass:full-smoothness,ass:bounded-variance-zeroth-order}, the query complexity to the stochastic evaluation oracle is $\tilde{\Ocal}\left(\frac{d^{2}L^2\sigma}{\alpha^{5/2}\epsilon^{3/2}}\right)$, or under \cref{ass:smoothness,ass:bounded-variance-zeroth-order} the query complexity to the stochastic evaluation oracle is $\tilde{\Ocal}\left(\frac{d^{3/2}L^2\sigma}{\alpha^{5/2}\epsilon^{3/2}}\right)$.
    
\end{restatable}
\begin{proof}
  By \cref{lem:LSI-one-step-convergence}, one-step equation can be written as 
  \begin{align}
  \KL(\mu_{k+1}||\pi) &\leq   e^{-3\alpha \eta/2} \left[\left(1+\frac{32\eta^3L^4}{\alpha}\right)\KL(\mu_k||\pi)+6\eta \hat{\sigma}^2 +16\eta^2 dL^2\right]\\
  &\leq  e^{-\alpha \eta} \KL(\mu_k||\pi)+6\eta \hat{\sigma}^2 +16\eta^2 dL^2.
\end{align}
Since for $\eta \leq \frac{\alpha}{8L^2}$, $1+\frac{32\eta^3L^4}{\alpha} \leq 1+ \frac{\alpha \eta}{2}\leq e^{\alpha \eta/2}$.
Unrolling the recursion, we have
  \begin{align}
  \KL(\mu_{k}||\pi) &\leq  e^{-\alpha \eta k} \KL(\mu_0||\pi)+\frac{6\eta \hat{\sigma}^2 +16\eta^2 dL^2}{1-e^{-\alpha \eta }}\\
  &\leq  e^{-\alpha \eta k} \KL(\mu_0||\pi)+\frac{8 \hat{\sigma}^2 +32\eta dL^2}{\alpha}\\
  &\leq  e^{-\alpha \eta k} \KL(\mu_0||\pi)+\frac{8 \hat{\sigma}^2 +32\eta dL^2}{\alpha}.
  \end{align}
The second inequality is due to the fact that for $\eta\leq \frac{\alpha}{8L^2}$, $1-e^{-\alpha \eta}\geq \frac{3}{4}\alpha \eta$ when $\alpha \eta \leq \frac{1}{4}$. We set $\eta \leq \frac{\epsilon \alpha}{128dL^2}$ and $\hat{\sigma}^2\leq \frac{\alpha\epsilon}{32}$ and $k\geq \frac{1}{\alpha \eta} \log\left(\frac{2\KL(\mu_0||\pi)}{\epsilon} \right)$ so that $\KL(\mu_k||\pi)\leq \epsilon$. The number of calls to the stochastic evaluation oracle under \cref{ass:full-smoothness,ass:bounded-variance-zeroth-order} to achieve $\hat{\sigma}^2\leq \frac{\alpha\epsilon}{32}$ at each iteration is $\Ocal\left(\frac{d\sigma}{\alpha^{1/2} \epsilon^{1/2}}\right)$ by \cref{thm:zeroth-order-gradient1}. Hence, the total number of calls to the stochastic evaluation oracle is $\tilde{\Ocal}\left(\frac{d^{2}L^2\sigma}{\alpha^{5/2}\epsilon^{3/2}}\right)$. Similarly, under \cref{ass:smoothness,ass:bounded-variance-zeroth-order} the number of calls to stochastic evaluation at each iteration is $\Ocal\left(\frac{d^{1/2}\sigma}{\alpha^{1/2} \epsilon^{1/2}}\right)$ by \cref{thm:smooth-gradient}. Hence, the total number of calls to the stochastic evaluation oracle is $\tilde{\Ocal}\left(\frac{d^{3/2}L^2\sigma}{\alpha^{5/2}\epsilon^{3/2}}\right)$.
\end{proof}

This result holds for several distance metrics. Compared with classical results, \cite{roy2021stochasticzerothorderdiscretizationslangevin} analyzes zeroth-order LMC under \cref{ass:full-smoothness,ass:bounded-variance-zeroth-order} and obtains evaluation complexity \(\Ocal(d^3\sigma^2/\epsilon^4)\) for convergence in \(\W_2\) (Theorem~3.2 of~\cite{roy2021stochasticzerothorderdiscretizationslangevin}). Our algorithm uses $\Tilde{\Ocal}\frac{d^2\sigma}{\epsilon^{3/2}})$ evaluation queries under the same assumptions, yielding polynomial improvements in \(d\), \(\sigma\), and \(\epsilon\).

\section{Applications to Optimization}
\label{sec:application}
\subsection{Finite-Sum Non-convex Optimization}
\label{sec:finite-sum-optimization}
In this section, we derive a finite-sum optimization algorithm by running our sampler at low temperature, equivalently by scaling \(f\) by a large inverse-temperature parameter \(\beta\). The following theorem quantifies how large \(\beta\) should be to find an approximate optimizer of \(f\).

\begin{restatable}{lemma}{optimizerbetaone}
\label{lem:optimizerbeta1}
Let $\pi^{\beta} = \frac{e^{-\beta f(\x)}}{\int e^{-\beta f(\x)}d\x}$. Suppose that $f$ satisfies \cref{assumption:LSI}. Then, if $\beta = \Theta(d/\epsilon)$, sampling from $\pi^{\beta}$ returns a point $\tilde{x}$ such that $\mathbb{E}[f(\tilde{x})-f(x^{\star})]\leq \epsilon$.
\end{restatable}
\begin{proof}
Let $f(x^{\star})=\sup_{\x\in \mathbb{R}^d} f = 0$ without loss of generality. 
Consider the following quantity that quantifies the optimization error,
\begin{equation}
    \varepsilon(\beta) = \mathbb{E}_{\pi^{\beta}}[f(\x)],
\end{equation}
which can be written explicitly,
\begin{align}
    \varepsilon(\beta) = \frac{\int_{\mathbb{R}^d}f(\x)\exp(-\beta f(\x))d\x }{\int_{\mathbb{R}^d}\exp(-\beta f(\x))d\x}.
\end{align}
Consider the derivative of $\varepsilon(\beta)$ with respect to $\beta$.
\begin{align}
    \frac{d\varepsilon(\beta)}{d\beta} &= \frac{\beta \left(\int_{\mathbb{R}^d}f(\x)\exp(-\beta f(\x))d\x\right)^2  -\beta\left(\int_{\mathbb{R}^d}f(\x)\exp(-\beta f^2(\x))d\x\right)\left(\int_{\mathbb{R}^d}\exp(-\beta f(\x))\right)d\x)}{\left(\int_{\mathbb{R}^d}\exp(-\beta f(\x))d\x\right)^2}\\
    & =  -\text{Var}(\beta^{1/2}f(\x)).
\end{align}
Since variance is positive quantity, $\frac{d\varepsilon(\beta)}{d\beta}$ is negative and its absolute value is bounded by
\begin{align}
    \left|\frac{d\varepsilon(\beta)}{d\beta} \right|\leq \text{Var}(\beta^{1/2}(f(\x)))\leq M\beta \text{Var}(\x)\leq \Ocal(d),
\end{align}
where the last inequality is due to the fact that the tails of $\pi^{\beta}$ are upper bounded by a Gaussian with variance $\Omega(1/\beta)$ due to the Log-Sobolev inequality. Since $\varepsilon(\infty)=0$, we can write 
\begin{align}
    \varepsilon(\beta) = \lim_{\beta_u\to \infty}\int_{\beta_{u}}^{\beta} \frac{d\varepsilon(\beta)}{d\beta }d\beta \leq \Ocal(d/\beta).
\end{align}
Then, for some $\beta>\Theta(d/\epsilon)$, $\mathbb{E}_{\pi^{\beta}}[f(\x)-f(\x^{\star})]=\varepsilon(\beta)\leq \epsilon$. This concludes the proof.
\end{proof}

\begin{theorem}[Quantum Finite-sum Optimization Upper Bound]
\label{thm:quantum-finite-sum-upper}
    Suppose $f$ is a $\mu$-strongly convex function. Then, there exists a quantum algorithm that outputs a point $\widetilde{\x}$ such that $\mathbb{E}[f(\widetilde{\x})-f(\x^{\star})]\leq \epsilon$ using $\Tilde{\Ocal}\left(\frac{L^2d}{\mu^2}\left(nd^{1/2} + \frac{d^{3/2}n^{1/3}}{\epsilon}\right)\right)$ queries to the quantum stochastic gradient oracle.
\end{theorem}
\begin{proof}
    Suppose that we sample from the distribution $\pi_{\beta}(\x)\propto \exp(-\beta f(\x))$. 
    Since we can only approximately sample from $\pi_{\beta}$, we need to take the sampling error into account as well. Suppose the quantum sampling algorithm in~\cref{thm:qlmc} returns a sample from $\rho$ instead of $\pi_{\beta}$. From Corollary C.1.1 in~\cite{NEURIPS2022_78e839f9}, the sufficient conditions for a sample from $\rho$ to give $\epsilon$ close approximation to the optimum in expectation are (i) $LW_{2}^2 (\rho, \pi)\leq \epsilon/2$ and (ii) $\mathbb{E}_{x\sim\pi}[f(x)-f(x^{\star})]\leq \epsilon/2$ where $W_{2}$ is the Wasserstein distance. By \cref{lem:optimizerbeta1}, for large $\beta \geq c\frac{d}{\epsilon}$, we have $\mathbb{E}_{x\sim \pi_{\beta}}[f(\widetilde{\x})-f(x^{\star}]\leq \epsilon$. Therefore, the latter condition is satisfied. Then, all we need is to characterize the sampling complexity of $\pi_{\beta}$ in terms of closeness in Wasserstein distance. Since $f$ is strongly convex, both LSI constant and smoothness constants scale linearly. Therefore, by~\cref{thm:qlmc} the sampling complexity is $\Tilde{\Ocal}\left(\frac{L^2\log(\KL(\mu_0||\pi))}{\mu^2}\left(nd^{1/2} + \frac{Ld^{3/2}n^{1/3}}{\mu\epsilon}\right) \right)$. Suppose that we choose the initial distribution to be a Gaussian $\mathcal{N}\left(0, \frac{1}{L}I\right)$. Then, by Lemma 1 in~\cite{NEURIPS2019_65a99bb7}, $\log(\mu_0||\pi) = \Ocal(d)$. This concludes the proof.
\end{proof}

This matches the upper bound of~\cite{pmlr-v235-zhang24bz} in its \(n\)-dependence. The remaining parameter dependencies are suboptimal; this is expected because, in the strongly convex setting, sampling is not the most efficient optimization method in terms of parameters such as \(L\) and \(\mu\). We have simplified the proof by assuming strong convexity, but one can also impose additional conditions, such as the Morse-type conditions used in~\cite{NEURIPS2022_78e839f9}, to obtain a quantum upper bound of the form \(\Ocal(n+n^{1/3}\epsilon^{-1})\) for a broader class of non-convex functions satisfying an LSI.

\subsection{Non-smooth Approximately Convex Optimization}
\label{sec:nonsmooth-optimization}
Although the previous sampling results assume smoothness (Lipschitz gradients), they also yield algorithms for non-smooth optimization. We consider objectives satisfying the following approximate-convexity and Lipschitz continuity assumptions.

\begin{assumption}[Approximate Convexity]
\label{ass:approx-convex}
Let $f$ be a differentiable function, we say that $f:\mathbb{R}^d\to \mathbb{R} $ is an $\epsilon$-approximately convex function, if there exists a strongly convex function $F$ such that for all $\x$,
\begin{equation}\label{eq:approx-convex}
    |F(\x) - f(\x)|\leq \frac{\epsilon}{d}.
\end{equation}
\end{assumption}

\begin{assumption}[Lipschitz Continuity]\label{ass:Lipschitz}
    For all $\x, \y \in \mathbb{R}^d$, $f:\mathbb{R}^d\to \mathbb{R}$ satisfies,
    \begin{align}
        |f(\x) -f(\y)|\leq M\|\x-\y\|.
    \end{align}
\end{assumption}

The goal is to find an approximate minimizer $\x^{\star}$ such that
$
    |f(\x^{\star}) - \min_{\x}f(\x)|\leq \epsilon.
$
This setting arises naturally in empirical risk minimization (ERM), where the objective is defined through empirical observations. Even if the population objective is smooth and strongly convex, subsampling can make the empirical objective non-smooth and non-convex. Similar settings also arise in the study of escaping local minima, both classically~\cite{pmlr-v40-Belloni15} and quantumly~\cite{10.5555/3600270.3600498}, under access to a stochastic evaluation oracle. Our approach differs from previous techniques by sampling from a Gibbs distribution associated with a smooth approximation of \(f\).

Since $f$ is not smooth, we consider the smoothed approximation
\begin{align}
        f_v(\x) = \mathbb{E}_{u\sim \mathcal{B}_d(0, 1) }[f(\x+v\uu)],
\end{align}
where $\mathcal{B}$ is the unit $L_2$ ball around the center. The local properties of $f_v$ are known and given by the following proposition. 

\begin{proposition}
\label{prop:local-properties}
    If $f$ satisfies \cref{ass:Lipschitz}, then $f_v$ satisfies $|f_v(\cdot) - f(\cdot)|\leq vM$ \quad and\quad  $|\nabla f_v(\x) - \nabla f_v(\y)|\leq L \|\x-\y\|$,
       where $L = c M \sqrt{d}v^{-1}$ for some constant $c>0$.
\end{proposition}

To characterize the runtime of the optimization algorithm, we use the following LSI perturbation lemma of Holley--Stroock~\cite{Holley1987LogarithmicSI}.
\begin{lemma}\label{lem:Halley}
Let $\rho$ be the Log-Sobolev constant of the Gibbs distribution with potential $F$. Then, the Log-Sobolev constant of $f$ satisfies
\begin{align}
    \alpha\geq \rho e^{-|\sup_x(f(x) - F(x)) - \inf_x (f(x)-F(x))|}.
\end{align}
\end{lemma}

First, note that \(\mathbb{E}_{\uu}\|\nabla f(\x+v\uu)-\nabla f_v(\x)\|^2\leq 4M^2\), since \(\|\nabla f(\x)\|\leq M\) by Lipschitz continuity.

The essential idea is that the Gibbs distribution localizes around the global minimizers of \(f\) as \(\beta\) increases. Another challenge in sampling from \(f_v\) is that the log-Sobolev constant may depend on the dimension. Using the Holley-Stroock LSI perturbation result~\cite{Holley1987LogarithmicSI}, we show how this dimension dependence enters the total evaluation complexity. We now state the main result.

\begin{restatable}{lemma}{optimizerbeta}
\label{lem:optimizerbeta}
Let $\pi_{v}^{\beta} = \frac{e^{-\beta f_v(\x)}}{\int e^{-\beta f_v(\x)}d\x}$. If $\beta = \Theta(d/\epsilon)$ and $v\leq \Ocal(\frac{\epsilon}{Md})$, then sampling from $\pi_v^{\beta}$ returns an $\epsilon$-approximate optimizer for $f$ with probability at least $0.1$.
\end{restatable}
\begin{proof}
Consider the following quantity that quantifies the optimization error with respect to the original distribution $\pi^{\beta}$.
\begin{equation}
    \varepsilon(\beta) = \mathbb{E}_{\pi^{\beta}}[f(\x)],
\end{equation}
which can be written explicitly,
\begin{align}
    \varepsilon(\beta) = \frac{\int_{\mathbb{R}^d}f(\x)\exp(-\beta f(\x))d\x }{\int_{\mathbb{R}^d}\exp(-\beta f(\x))d\x}.
\end{align}
Consider the derivative of $\varepsilon(\beta)$ with respect to $\beta$.
\begin{align}
    \frac{d\varepsilon(\beta)}{d\beta} &= \frac{\beta \left(\int_{\mathbb{R}^d}f(\x)\exp(-\beta f(\x))d\x\right)^2  -\beta\left(\int_{\mathbb{R}^d}f(\x)\exp(-\beta f^2(\x))d\x\right)\left(\int_{\mathbb{R}^d}\exp(-\beta f(\x))\right)d\x)}{\left(\int_{\mathbb{R}^d}\exp(-\beta f(\x))d\x\right)^2}\\
    & =  -\text{Var}(\beta^{1/2}f(\x))).
\end{align}
Since variance is positive quantity, $\frac{d\varepsilon(\beta)}{d\beta}$ is negative and its absolute value is bounded by
\begin{align}
    \left|\frac{d\varepsilon(\beta)}{d\beta} \right|\leq \text{Var}(\beta^{1/2}(f(\x)))\leq M\beta \text{Var}(\x)\leq \Ocal(d),
\end{align}
where the last inequality is due to the fact that tails of $\pi^{\beta}$ is upper bounded by a Gaussian with variance $\Omega(1/\beta)$. Since $\varepsilon(\infty)=0$, we can write 
\begin{align}
    \varepsilon(\beta) = \lim_{\beta_u\to \infty}\int_{\beta_{u}}^{\beta} \frac{d\varepsilon(\beta)}{d\beta }d\beta \leq \Ocal(d/\beta).
\end{align}
Let $f(x^{\star})=\min_\x f = 0$ without loss of generality. Then, for $\beta>\Theta(d/\epsilon)$, $\mathbb{E}_{\pi^{\beta}}[f(\x)-f(\x^{\star})]=\varepsilon(\beta)\leq \epsilon$. By Markov's inequality
\begin{equation}
    \Pr[f(\x)-f(\x^{\star})\geq \epsilon)]\leq 0.05. 
\end{equation}
For $\nu\leq \Ocal(\frac{\epsilon}{Md})$,
\begin{equation}
   \beta(f_\nu(\x)-f(\x))\leq \beta M\nu \leq \Ocal(1).
\end{equation}
Therefore, the total variation distance between $\pi^{\beta}$ and $\pi_\nu^{\beta}$ can be made smaller than $0.05$ by scaling $\nu$ by a constant. Then,
\begin{equation}
        \Pr[f(\x)-f(\x^{\star})\geq \epsilon)]\leq 0.1. 
\end{equation}
This concludes the proof.

\end{proof}

\begin{restatable}{theorem}{optimization}
\label{thm:optimization}
    Suppose that $f$ satisfies \cref{ass:approx-convex,ass:Lipschitz}. Then,
    there exists a quantum algorithm that returns $\epsilon$ approximate minimizer for $f$ with high probability using $\Tilde{\mathcal{O}}\left(\frac{d^{9/2}}{\epsilon^{3/2}}\right)$ queries to the stochastic evaluation oracle for $f$.
\end{restatable}
\begin{proof}
    We consider the potential function $\beta f_v(x)$ where $\beta$ is the inverse temperature parameter. By \cref{lem:optimizerbeta}, sampling from $\pi_v^{\beta}\propto e^{-\beta f_v }$ returns $\frac{\epsilon}{2}$ approximate minimizer for $f$ with high probability (say $0.9$) for sufficiently large $\beta = \Ocal (\frac{d}{\epsilon})$. Suppose that we sample from a probability distribution $\mu$ such that 
    \begin{align}
    \TV(\mu,\pi_v^{\beta})\leq 0.1.
    \end{align}
    Then, the sample must be $\frac{\epsilon}{2}$ minimizer for $f$ with probability at least $0.8$. Therefore, it is sufficient to sample from $\pi_{v}^{\beta}$ up to a constant $\TV$ distance.

    We need to characterize the sampling complexity from $\pi_{v}^{\beta}$. From Bakry Emery theorem, Log Sobolev constant $\rho$ of $\beta F$ satisfies $\rho \geq \frac{\beta \mu}{2}$ where $\mu$ is the strong convexity constant of $F$. Let $M' = \max(M,1)$ and take $v = \frac{\epsilon}{2M'd}$. Then using LSI perturbation result \cref{lem:Halley} by Holley-Strock, we have $\alpha \geq \frac{\beta \mu}{2}e^{-3\beta \epsilon/d} = \Omega(\frac{\mu d}{\epsilon})$. The uniform perturbation holds due to the inequality $|f_v-F|\leq |f_v-f|+|f-F|\leq vM+\frac{\epsilon}{d}\leq \frac{3\epsilon}{2d}$. Since $\beta f_v$ is a smooth function with smoothness constant $L=\Ocal(\frac{\beta M\sqrt{d}}{v})=\Ocal(\frac{d^{5/2}M^2}{\epsilon^2})$ by \cref{prop:local-properties}, the number of calls to stochastic evaluation oracle to sample from $\pi_v$ is $\Tilde{\Ocal}(\frac{L^2d^2}{\alpha^{5/2}})=\Tilde{\Ocal}(\frac{M^4 d^{9/2}}{\mu^{5/2}\epsilon^{3/2}})$ by \cref{thm:qzlmc}. Hence, we can optimize $f$ in polynomial time. 
\end{proof}

The closest result to our setting is~\cite{10.5555/3600270.3600498}, whose stochastic query complexity is \(\Tilde{\Ocal}(d^5/\epsilon)\), although the assumptions are not directly comparable. The authors assume additive sub-Gaussian noise and convexity of \(F\) on a bounded domain, but not strong convexity. Because these assumptions differ from ours, the comparison should be interpreted cautiously. Under our assumptions, the algorithm improves the dimension dependence at the cost of worse \(\epsilon\)-dependence, a common tradeoff for sampling-based optimization methods.

We also note the work by Augustino et al.~\cite{augustino2025fastconvexoptimizationquantum} and Chakrabarti et al.~\cite{chakrabarti2025speedupsconvexoptimizationquantum}, which also investigate zeroth-order stochastic convex optimization under assumptions similar to those in~\cite{10.5555/3600270.3600498}. They propose algorithms with query complexities \(\widetilde{\Ocal}(d^{9/2}/\epsilon^7)\) and \(\widetilde{\Ocal}(d^3/\epsilon^5)\), respectively. Although both approaches have worse \(\epsilon\)-dependence than ours, their assumptions and problem settings differ significantly from ours.

\section{Acknowledgment}
Chunhao Wang and Guneykan Ozgul acknowledge support from National Science Foundation grant CCF-2238766 (CAREER). Xiantao Li acknowledges support from National Science Foundation grants DMS-2111221 and CCF-2312456.

\bibliographystyle{alpha}
\bibliography{bibo}

\addtocontents{toc}{\protect\setcounter{tocdepth}{1}}
\appendix

\section{Technical Proofs for Sampling}
\label{sec:HMC-appendix}
\subsection{Technical Proofs for HMC}
We start with the following result in \cite{pmlr-v139-zou21b} that quantifies the convergence of the stochastic gradient Hamiltonian Monte Carlo algorithm in Wasserstein distance.
\begin{theorem}[Theorem 4.4 in \cite{pmlr-v139-zou21b}]
\label{thm:HMC-convergence}
    Under \cref{assumption:strong-convexity,assumption:component-smoothness}, let $D= \|\x^{0} -\arg\min_{\x}(f(\x))\|$ and $\mu_T$ be the distribution of the iterate $\x^{T}$, then if the step size satisfies $\eta = O(L^{1/2}\sigma^{-2}\kappa^{-1} \wedge 
 L^{-1/2}) $ and $K = 1/(4\sqrt{L}\eta)$, the output of HMC satisfies
 \begin{equation}
 \label{eq:HMC-convergence}
     \W_2(\mu_T, \pi)\leq (1-(128\kappa)^{-1})^{\frac{T}{2}}(2D+ 2d/\mu)^{1/2} + \Gamma_1\eta^{1/2}+\Gamma_2\eta,
 \end{equation}
 where $\Gamma_1^2 = O(
L^{-3/2}\sigma^2\kappa^2)$ and $\Gamma_2^2 = O(\kappa^2(LD+\kappa d + L^{-1/2}\sigma^2\eta))$ where $\sigma^2 =\max_{t\leq T}\mathbb{E}\|\g(\x_k,\boldsymbol{\xi}_k)-\nabla f(\x_k)\|^2$ is the upper bound on the variance of the gradients in the trajectory of \algname{SG-HMC} algorithm. 
\end{theorem}
This is a generic result that applies to any HMC algorithm under \cref{assumption:strong-convexity,assumption:component-smoothness} that uses stochastic gradients with variance upper bounded by $\sigma^2$. Note that we do not assume a uniform upper bound for $\sigma$ that is independent of problem parameters. Instead, the variance upper bound depends on the trajectory of the algorithm, which can be characterized using theoretical analysis.

\label{sec:proof-qsvrg-hmc}
\begin{lemma} 
    \label{lem:svrg-variance1}
    Under \cref{assumption:component-smoothness}, if the initial point satisfies $\|\x^{0}-\x^{\star}\|\leq \frac{d}{\mu}$, then it holds that
    \begin{equation}
        \mathbb{E}_i\|\g (\x_k,\xi) - \nabla f(\x_k)\|^2\leq L^2 \|\x_k - \widetilde{\x}\|^2,
    \end{equation}
    where $\widetilde{\x}=\x_{k'<k}$ is the last iteration the full gradient is computed.
\end{lemma}
\begin{proof}
The proof simply follows from the definition of variance in the SVRG algorithm and the smoothness of each component.
    \begin{align}
          \mathbb{E}_i\|\g_i (\x_k,\xi) - \nabla f(\x_k)\|^2&\leq \mathbb{E}_i\|\nabla f_i(\x_k)-\nabla f_i(\widetilde{\x})+\nabla f(\widetilde{\x})-\nabla f (\x_k)  \|^2\\
          &\leq \mathbb{E}_i\|\nabla f_i(\x_k)-\nabla f_i(\widetilde{\x})\|^2\\
          &\leq L^2 \|\x_k - \widetilde{\x}\|^2.
    \end{align}
\end{proof}
Lemma \ref{lem:svrg-variance1} allows us to set the target variance in quantum mean estimation to be $L^2\|\x_k-\widetilde{\x}\|/b^2$. Hence, each mean estimation call takes $\Ocal(d^{1/2}b)$ gradient evaluations by \cref{lem:var-reduction}. The following lemma characterizes the variance of the stochastic gradients along the trajectory of the algorithm.

\begin{lemma}[Modified Lemma C.2 in \cite{pmlr-v139-zou21b}]
\label{lem:modified-variance-lemma}
    Let $\g(\x_k, \boldsymbol{\xi}_k)$ be the vector computed using the unbiased quantum mean estimation algorithm in \algname{QHMC-SVRG}. Then, under \cref{assumption:component-smoothness},
    \begin{equation}
          \mathbb{E}\|\g(\x_k, \boldsymbol{\xi}_k) - \nabla f(\x_k)\|^2\leq  \frac{768 m^2 L^2 \eta^2\kappa d}{b^2},
    \end{equation}
    where the expectation is over both the iterate $\x_k$ and the noise in quantum mean estimation $\boldsymbol{\xi}_k$.
\end{lemma}

\label{sec:proof-qcv-hmc}
\begin{lemma}[Modified Lemma C.4 in \cite{pmlr-v139-zou21b}]
\label{lem:modified-variance-lemma2}
    Let $\g(\x_k, \boldsymbol{\xi}_k)$ be the vector computed using the unbiased quantum mean estimation algorithm in \algname{QHMC-CV}. Then, under \cref{assumption:component-smoothness},
    \[
    \mathbb{E}\|\g(\x_k, \boldsymbol{\xi}_k) - \nabla f(\x_k)\|^2\leq  \frac{688Ld\kappa}{b^2},
    \]
    where the expectation is over both the iterate $\x_k$ and the noise in quantum mean estimation $\boldsymbol{\xi}_k$.
\end{lemma}
\subsection{Technical Proofs for QSVRG-LMC}
\label{sec:proof-qsvrg-lmc}
We start with the following lemma that characterizes the variance of the quantum stochastic gradients in \algname{QSVRG-LMC} in terms of the distance between the current iterate and the reference point where the full gradient is computed.
\begin{restatable}{lemma}{svrgvariancelemma}
\label{lem:svrg-quantum-reduce}
    Let $\tilde{\x}$ be any iteration where $\algname{QSVRG-LMC}$  computes the full gradient. Then under \cref{assumption:component-smoothness}, the quantum stochastic gradient $\g_k$ at $\x_k$ 
    that is computed using $\tilde{\x}$ as a reference point in $\algname{QSVRG-LMC}$ satisfies
    \begin{equation}
        \mathbb{E}[\|\g_k-\nabla f(\x_k)\|^2]\leq \frac{L^2 \|\x_k-\tilde{\x}\|^2}{b^2}
    \end{equation}
    using $\Tilde{O}(d^{1/2} b)$ gradient computations.
\end{restatable}

\begin{proof}
    Recall that SVRG computes the stochastic gradient $\Tilde{\g}$ at $\x_k$ by the following.
\begin{align}
    \Tilde{\g}_k = \nabla f_i(\x_k) - \nabla f_i(\tilde{\x})+\nabla f(\tilde{\x}),
\end{align}
where $\tilde{\x}$ is the last iteration the full gradient is computed and $i$ is a component randomly chosen from $[n]$. Let $\sigma_k^2 = \mathbb{E}\|\Tilde{\g}_k - \nabla f(\x_k) \|^2$. Then, $\sigma_k^2$ can be bounded in terms of the distance between $\x_k$ and $\tilde{\x}$.
\begin{align}
    \sigma_k^2&=\mathbb{E}[\|\nabla f_i(\x_k)-\nabla f_i(\Tilde{\x}) +\nabla f(\Tilde{\x})-\nabla f(\x_k)  \|^2]\\
    &= \mathbb{E}[\|\nabla f_i(\x_k)-\nabla f_i(\Tilde{\x})\|^2] -\left(\mathbb{E}[\nabla f_i(\x_k)-\nabla f_i(\Tilde{\x})]\right)^2\\
    &  \leq \mathbb{E} [\|\nabla f_i(\x_k)-\nabla f_i(\Tilde{\x}) \|^2]\\
      &\leq L^2 \|\x_k - \Tilde{\x} \|^2,
\end{align}
where the equality follows from the fact that $\nabla f_i$ is an unbiased estimator for $\nabla f$ and the last line follows from \cref{assumption:component-smoothness}. Hence, using unbiased quantum mean estimation in \cref{lem:var-reduction}, we can obtain a random vector $\g_k$ such that,
\begin{align}
      \mathbb{E}\|\g_k-\nabla f(\x_k)\|^2\leq \frac{L^2 \|\x_k-\tilde{\x}\|^2}{b^2}
\end{align}
by using $\Tilde{O}(d^{1/2} b)$ calls to the gradient oracle.
\end{proof}
To be able to apply \cref{lem:LSI-one-step-convergence}, we need to characterize the expected upper bound on the variance of the stochastic gradients over the algorithm trajectory for SVRG. 
\svrgvariance*
\begin{proof}
Let $\Tilde{\x} = \x_{k'}$. Then, by \cref{lem:svrg-quantum-reduce}, quantum stochastic gradient $\g_k$ satisfies
\begin{align}
   \label{eq:max-var}
     \mathbb{E}[\|\g_k-\nabla f(\x_k)\|^2]\leq \frac{L^2 \mathbb{E}\|\x_k-\tilde{\x}\|^2}{b^2}.
\end{align}
Let $\tilde{\x} = \y_0 $ and $\x_k = \y_k$, then using the update rule of Langevin Monte Carlo,
\begin{align}
    \mathbb{E}[\|\x_k -\Tilde{\x}\|^2] &= \mathbb{E}\left[\left\|\sum_{r=1}^l(\y_{r} - \y_{r-1})\right\|^2\right] = \mathbb{E}\left[\left\|\sum_{r=1}^l- \eta \g_{r-1} +\sqrt{2\eta}\boldsymbol{\epsilon}_{r-1})\right\|^2\right] \\
    &\leq \mathbb{E}\left[2\eta^2\left\|\sum_{r=1}^l\g_{r-1} \right \|^2 +4\eta \left\|\sum_{r=1}^l\boldsymbol{\epsilon}_{r-1} \right\|^2   \right]\\
    &\leq 2\eta^2 m\sum_{r=1}^l\mathbb{E}\left\| \g_{r-1}\right\|^2 + 4\eta \sum_{r=1}^l\|\boldsymbol{\epsilon}_{r-1}\|^2\\
    &\leq 2\eta^2 m\sum_{r=1}^l\mathbb{E}\left\| \g_{r-1}\right\|^2 + 4\eta d m.
\end{align}
The first inequality is due to Young's inequality and the second inequality follows from the fact that the Gaussian noises at different iterations are independent and the fact that $l\leq m$. Defining $\sigma_{\textrm{max}}^2= \max_k\mathbb{E}\|\sigma_k\|^2$, we can write the first term on the right-hand side in terms of $\sigma_{\textrm{max}}^2$,
\begin{align}
    \mathbb{E}[\|\g_{r}\|^2]  &= \mathbb{E}\|\g_r - \nabla f(\x_r) +\nabla f (\x_r) \|^2 \\
    &\leq 2  \mathbb{E}\|\g_r - \nabla f(\x_r) \|^2 + 2\|\nabla f (\x_r) \|^2\\
    &\leq 2 \sigma_{\textrm{max}}^2 + \frac{8L^2}{\alpha}\KL(\mu_r||\pi) + 4dL,
\end{align}
and using \cref{eq:max-var},
\begin{align}
      \sigma_{\textrm{max}}^2\leq  \frac{4L^2m^2 \eta^2 \sigma_{\textrm{max}}^2}{b^2} +\frac{16L^4 \eta^2m }{b^2 \alpha}\sum_{r=1}^{l}\KL(\mu_{r-1}||\pi) +\frac{8dL^3\eta^2m^2}{b^2}+\frac{4\eta dmL^2}{b^2}.
\end{align}
If we set $\eta^2 \leq \frac{1}{6 L^2 m^2 }$, we obtain
\begin{align}
   \sigma_{k'+l}^2\leq \frac{32L^4 \eta^2 m}{b^2\alpha}\sum_{r=1}^{l} \KL(\mu_{r-1}||\pi) + \frac{8\eta d m L^2}{b^2}.
\end{align}
\end{proof}

\qsvrgconvergence*
\begin{proof}
Let $l<k$ be the last iteration the full gradient is computed. Then, using \cref{lem:LSI-one-step-convergence,lem:svrgvariance}, we can write one step bound as follows.
\begin{align}
    \label{eq:one-step-bound}
    \KL(\mu_{k+1}||\pi) &\leq   e^{-3\alpha \eta/2} \left[\left(1+\frac{32\eta^3L^4}{\alpha}\right)\KL(\mu_k||\pi)+\frac{192 m \eta^3  L^4}{b^2\alpha} \sum_{r=l}^{k} \KL(\mu_{r}||\pi) +\frac{48m \eta^2 dL^2}{b^2}+16\eta^2 d L^2\right].
\end{align}  
First, we claim that the following inequality is true.
\begin{align}
    \label{eq:svrg-claim}
     \KL(\mu_{k+1}||\pi)\leq e^{-\alpha \eta k}\KL(\mu_0\|\pi)+\frac{48m \eta^2 dL^2+16\eta^2 d L^2b^2}{b^2(1-e^{-\alpha \eta})}.
\end{align}
To prove \cref{eq:svrg-claim}, we use induction. For $k=1$, the statement holds due to \cref{eq:one-step-bound}. That is,
\begin{align}
    \KL(\mu_{1}||\pi) &\leq   e^{-3\alpha \eta/2} \left[\left(1+\frac{224\eta^3L^4}{\alpha}\right)\KL(\mu_0||\pi)+\frac{48m \eta^2 dL^2}{b^2}+ 16\eta^2 d L^2\right] \\
    &\leq  e^{-\alpha \eta}\KL(\mu_0\|\pi)+\frac{48m \eta^2 dL^2}{b^2}+ 16\eta^2 d L^2\\
    &\leq e^{-\alpha \eta}\KL(\mu_0\|\pi)+\frac{48m \eta^2 dL^2+16\eta^2 d L^2b^2}{b^2(1-e^{-\alpha \eta})}.
\end{align}  
The first inequality is due to the fact that $m\leq b^2$. The second inequality holds since $\left(1+\frac{224\eta^3L^4}{\alpha}\right)\leq \left(1+ \frac{\eta \alpha }{2}\right)\leq e^{\alpha \eta / 2}$ since $\eta \leq \frac{\alpha}{24L^2 m}$. The third inequality follows from the fact that $1-e^{-\alpha \eta}\leq 1$. Next, assume that the statement holds for $k-1$, and then we prove the $k$-th step of induction.
\begin{align}
    \KL(\mu_{k}||\pi) &\leq   e^{-3\alpha \eta/2} \left[\left(1+\frac{32\eta^3L^4}{\alpha}\right)\KL(\mu_{k-1}||\pi)+\frac{192 \eta^3  L^4}{\alpha} \sum_{r=\ell}^{k-1} \KL(\mu_{\ell}||\pi) + \frac{48m \eta^2 dL^2+16\eta^2 d L^2b^2}{b^2}\right]\\
    &\leq e^{-3\alpha \eta/2}\left(1+\frac{32\eta^3L^4}{\alpha}\right)
    \left(e^{-\alpha \eta (k-1)}\KL(\mu_0\|\pi)+\frac{48m \eta^2 dL^2+16\eta^2 d L^2b^2}{b^2(1-e^{-\alpha \eta})}\right)\\
    &+e^{-3\alpha \eta/2}\frac{192 \eta^3  L^4}{\alpha} \sum_{r=l}^{k-1} \left(e^{-\alpha \eta r}\KL(\mu_0\|\pi)+\frac{48m \eta^2 dL^2+16\eta^2 d L^2b^2}{b^2(1-e^{-\alpha \eta})}\right) + \frac{48m \eta^2 dL^2+16\eta^2 d L^2b^2}{b^2}\\
    &\leq e^{-3\alpha \eta/2}\left(1+\frac{32\eta^3L^4}{\alpha}\right)
    \left(e^{-\alpha \eta (k-1)}\KL(\mu_0\|\pi)+\frac{48m \eta^2 dL^2+16\eta^2 d L^2b^2}{b^2(1-e^{-\alpha \eta})}\right)\\
    &+e^{-3\alpha \eta/2}\frac{192 m\eta^3  L^4}{\alpha} e^{m \alpha \eta }\left(e^{-\alpha \eta (k-1)}\KL(\mu_0\|\pi)+\frac{48m \eta^2 dL^2+16\eta^2 d L^2b^2}{b^2(1-e^{-\alpha \eta})}\right) + \frac{48m \eta^2 dL^2+16\eta^2 d L^2b^2}{b^2}\\
    &\leq e^{-3\alpha \eta/2}\left(1+\frac{32\eta^3L^4}{\alpha}\right)
    \left(e^{-\alpha \eta (k-1)}\KL(\mu_0\|\pi)+\frac{48m \eta^2 dL^2+16\eta^2 d L^2b^2}{b^2(1-e^{-\alpha \eta})}\right)\\
    &+e^{-3\alpha \eta/2}\frac{96 m\eta^3  L^4}{\alpha}\left(e^{-\alpha \eta (k-1)}\KL(\mu_0\|\pi)+\frac{48m \eta^2 dL^2+16\eta^2 d L^2b^2}{b^2(1-e^{-\alpha \eta})}\right) +\frac{48m \eta^2 dL^2+16\eta^2 d L^2b^2}{b^2}\\
    &\leq e^{-3\alpha \eta/2}\left(1+\frac{128\eta^3L^4}{\alpha}\right)
    \left(e^{-\alpha \eta (k-1)}\KL(\mu_0\|\pi)+\frac{48m \eta^2 dL^2+16\eta^2 d L^2b^2}{b^2(1-e^{-\alpha \eta})}\right)+ \frac{48m \eta^2 dL^2+16\eta^2 d L^2b^2}{b^2}\\
    &\leq e^{-\alpha \eta}
    \left(e^{-\alpha \eta (k-1)}\KL(\mu_0\|\pi)+\frac{48m \eta^2 dL^2+16\eta^2 d L^2b^2}{b^2(1-e^{-\alpha \eta})}\right)+\frac{48m \eta^2 dL^2+16\eta^2 d L^2b^2}{b^2}\\
    &\leq e^{-\alpha \eta k}\KL(\mu_0\|\pi)+\frac{48m \eta^2 dL^2+16\eta^2 d L^2b^2}{b^2(1-e^{-\alpha \eta})}\\
    &\leq  e^{-\alpha \eta k}\KL(\mu_0\|\pi)+\frac{64m \eta dL^2+24\eta d L^2b^2}{\alpha b^2}.
\end{align}  
The first two inequalities are due to \cref{eq:one-step-bound}. The third and fourth inequality follow from the fact that $k-l\leq m$ and $e^{m\alpha \eta}\leq  e^{\frac{\alpha^2}{8L^2}}\leq e^{\frac{1}{8}}\leq \frac{1}{2}$ for $\eta\leq \frac{\alpha}{8mL^2}$ and the fifth inequality holds since $\left(1+\frac{128\eta^3L^4}{\alpha}\right)\leq \left(1+ \frac{\eta \alpha }{2}\right)\leq e^{\alpha \eta / 2}$ for $\eta \leq \frac{\alpha}{24L^2 m}$. The final inequality follows from the fact that $1-e^{-\alpha \eta}\geq \frac{3}{4}\alpha \eta$ when $\alpha \eta \leq \frac{1}{4}$. This concludes the proof.
\end{proof}
\begin{lemma}[Stochastic-LMC One Step Convergence]
\label{lem:LSI-one-step-convergence}
   Let $\mu_k$ be the distribution of the iterate $\x_k$, then if the step size satisfies $\eta = \frac{2}{3\alpha} $,
\begin{align}
  \KL(\mu_{k+1}||\pi) \leq   e^{-3\alpha \eta/2} \left[\left(1+\frac{32\eta^3L^4}{\alpha}\right)\KL(\mu_k||\pi)+6\eta \sigma_{k}^2 +16\eta^2 dL^2\right],
\end{align}
 where $\sigma_{k}^2 = \mathbb{E}_{\x_k, \boldsymbol{\xi}_k}\|\g(\x_k, \boldsymbol{\xi}_k) - \nabla f(\x_k)\|^2 $.
\end{lemma}
\begin{proof}
We compare one step of LMC starting at $\x_k$ with stochastic gradients $\g(\x_k,\boldsymbol{\xi}_k)$ to the output of continuous Langevin SDE (\cref{eq:langevin-sde}) starting at $\x_k$ with true gradient $\nabla f(\x_t)$ after time $\eta$. This technique has been used to establish the convergence of the unadjusted Langevin algorithm with full gradients under isoperimetry by \cite{NEURIPS2019_65a99bb7}. We extend the analysis by \cite{NEURIPS2019_65a99bb7} to the stochastic gradient LMC. Assume that the initial point $\x_k$ and $\g(\x_k, \boldsymbol{\xi}_k)$ obey the joint distribution $\mu_0$. The randomness in $\g(\x_k, \boldsymbol{\xi}_k)$ depends both on the randomness in $\x_k$ and the randomness in the quantum mean estimation algorithm. Then, one step update of the LMC algorithm with stochastic gradient yields,
\[
\x_{k+1} = \x_k - \eta\g(\x_k, \boldsymbol{\xi}_k)+\sqrt{2\eta} \boldsymbol{\epsilon}_k.
\]
Alternatively, $\x_{k+1}$ can be written as the solution of the following SDE at time $t=\eta$,
\[
d\x_t = -\g_k dt +\sqrt{2}d\boldsymbol{W}_t
\]
where $\g_k = \g(\x_k, \boldsymbol{\xi}_k)$ and $\boldsymbol{W}_t$ is the standard Brownian motion starting at $\boldsymbol{W}_0 = 0$. Let $\mu_t(\x_k, \g_k, \x_t)$ be the joint distribution of $\x_k$, $\g_k$, and $\x_t$ at time $t$. Each expectation in the proof is over this joint distribution unless specified otherwise. 

Consider the following stochastic differential equation
\begin{equation*}
    d\boldsymbol{X} = \boldsymbol{v}(\boldsymbol{X})dt + \sqrt{2} d\boldsymbol{W},
\end{equation*}
where $\boldsymbol{v}$ is a smooth vector field and $\boldsymbol{W}$ is the Brownian motion with $\boldsymbol{W}_0 = 0$. The Fokker-Planck equation describes the evolution of the probability density function $\mu_t$ as follows:
\begin{equation}
\label{eq:fokker-planck}
    \frac{\partial \mu_t}{\partial t} = - \nabla \cdot (\mu_t \boldsymbol{v} ) + \Delta \mu_t,
\end{equation}
where $\nabla \cdot$ is the divergence operator and $\Delta$ is the Laplacian. Then, the Fokker Planck equation  gives the following evolution for the marginal density $\mu_t(\x|\x_k, \g_k) = \mu_t(\x_t=\x|\x_k, \g_k)$,
\begin{align}
   \frac{\partial \mu_t(\x|\x_k, \g_k)}{\partial t} =  \mathbf{\nabla }\cdot (\mu_t(\x|\x_k, \g_k) \g_k ) + \Delta \mu_t(\x|\x_k, \g_k).
\end{align}
Taking the expectation over both sides with respect to $(\x_k, \g_k) \sim \mu_0$,
\begin{align}
 \frac{\partial \mu_t(\x)}{\partial t} &= \mathbb{E}_{(\x_k, \g_k) \sim \mu_0} [\mathbf{\nabla }\cdot (\mu_t(\x|\x_k) \g_k )] +  \mathbb{E}_{(\x_k, \g_k) \sim \mu_0} [\Delta \mu_t(\x|\x_k)]   \\
 &=\int\limits_{\mathbb{R}^d}\mathbf{\nabla }\cdot (\mu_t(\x|\x_k, \g_k) \g_k )\mu_0(\x_k, \g_k)d\x_k d\g_k   + \int\limits_{\mathbb{R}^d}\Delta \mu_t(\x|\x_k, \g_k)\mu_0(\x_k, \g_k)d\x_k d\g_k   \\
  &=\int\limits_{\mathbb{R}^d}\mathbf{\nabla }\cdot (\mu_t(\x)\mu(\x_k, \g_k|\x_t = \x) \g_k) d\x_kd\g_k   +  \Delta \mu_t(\x)  \\
  &= \mathbf{\nabla }\cdot\left(\mu_t(\x)\mathbb{E}[\g_k-\nabla f(\x_k)|\x_t = \x]+  \mu_t(\x)\nabla \log \left( \frac{\mu_t(\x)}{\pi(\x)}\right) \right ).
\end{align}
Consider the time derivative of KL divergence between $\mu_t$ and $\pi$,
\begin{align}
\frac{d}{dt}\KL(\mu_t||\pi)&= \frac{d}{dt} \int\limits_{\mathbb{R}^d} \mu_t(\x)\log\left(\frac{\mu_t(\x)}{\pi(\x)}\right)  d\x \\
&=\int\limits_{\mathbb{R}^d} \frac{\partial\mu_t(\x)}{\partial t}\log\left(\frac{\mu_t(\x)}{\pi(\x)}\right)  d\x_t + \int\limits_{\mathbb{R}^d} \mu_t(\x)\frac{\partial}{\partial t}\log\left(\frac{\mu_t(\x)}{\pi(\x)}\right)d\x\\
&= \int\limits_{\mathbb{R}^d} \frac{\partial\mu_t(\x)}{\partial t}\log\left(\frac{\mu_t(\x)}{\pi(\x)}\right)  d\x_t + \int\limits_{\mathbb{R}^d}\frac{\partial \mu_t(\x)}{\partial t}d\x \\
&=  \int\limits_{\mathbb{R}^d} \frac{\partial\mu_t(\x)}{\partial t}\log\left(\frac{\mu_t(\x)}{\pi(\x)}\right)  d\x_t.
\end{align}
The last term in the third equality vanishes since $\mu_t$ is a probability distribution and its $L_1$ norm is always 1. Then the $\KL$ divergence evolves as
\begin{align}
    \frac{d}{dt}\KL(\mu_t||\pi) &= \int\limits_{\mathbb{R}^d} \mathbf{\nabla }\cdot\left(\mu_t(\x)\mathbb{E}[\g_k-\nabla f(\x)|\x_t = \x]+  \mu_t(\x)\nabla \log \left( \frac{\mu_t(\x)}{\pi(\x)}\right) \right )\log\left( \frac{\mu_t(x)}{\pi(\x)}\right)d\x\\
    &= -\int\limits_{\mathbb{R}^d} \mu_t(\x)\left\langle \mathbb{E}[\g_k-\nabla f(\x)|\x_t = \x]+  \nabla \log \left( \frac{\mu_t(\x)}{\pi(\x)}\right),  \nabla \log \left( \frac{\mu_t(\x)}{\pi(\x)}\right) \right \rangle d\x\\
    &= -\int\limits_{\mathbb{R}^d} \mu_t(\x) \left\| \nabla \log \left( \frac{\mu_t(\x)}{\pi(\x)}\right)\right\|^2d\x + \mathbb{E}\left\langle \nabla f(\x_t)-\g_k,  \nabla \log \left( \frac{\mu_t(\x)}{\pi(\x)}\right)\right\rangle. 
\end{align}
The second term can be bounded as follows:
\begin{align}
  \mathbb{E} \left\langle \nabla f(\x_t)-\g_k,  \nabla \log \left( \frac{\mu_t(\x)}{\pi(\x)}\right)\right\rangle &\leq  \mathbb{E} \left[\|\nabla f(\x_t)-\g_k \|^2 +\frac{1}{4}\left\|\nabla \log \left( \frac{\mu_t(\x)}{\pi(\x)}\right)\right\|^2\right]\\
  &=  \mathbb{E}\|\nabla f(\x_t)-\g_k\|^2 +\frac{1}{4} \FI(\mu_t||\pi)\\
  & =  \mathbb{E}\|\nabla f(\x_t)-\nabla f(\x_k) +\nabla f(\x_k)-\g_k\|^2+\frac{1}{4} \FI(\mu_t||\pi)\\
  &\leq  2\mathbb{E}\|\nabla f(\x_t)-\nabla f(\x_k)\|^2+2 \mathbb{E}_{\mu_t(\x_t,\x_k) }\|\nabla f(\x_k)-\g_k\|^2\\&+\frac{1}{4} \FI(\mu_t||\pi).
\end{align}
The first inequality holds since $\langle a,b \rangle \leq a^2 + \frac{b^2}{4}$. The last line follows from Young's inequality. Furthermore, using Lipschitzness of gradients of $f$, we have
\begin{align}
    \mathbb{E}\|\nabla f(\x_t)-\nabla f(\x_k)\|^2&\leq L^2  \mathbb{E}\|\x_t - \x_k\|^2\\
    &\leq L^2  \mathbb{E}\|-t\g_k+\sqrt{2t}\boldsymbol{\epsilon}_k \|^2\\
    &= t^2 L^2   \mathbb{E}_{\mu_0}\|\g_k\|^2+2tdL^2.
\end{align}
Plugging these back into the time derivative of $\KL$ divergence, we have
\begin{align}
      \frac{d}{dt}\KL(\mu_t||\pi)&\leq -\frac{3}{4}\FI(\mu_t||\pi)+ 2t^2L^2\mathbb{E}_{\mu_0}\|\g_k\|^2+ 2\mathbb{E}_{\mu_0}\|\nabla f(\x_k)-\g_k\|^2+4tdL^2\\
      &\leq -\frac{3}{4}\FI(\mu_t||\pi)+ (4t^2L^2 + 2)\mathbb{E}_{\mu_0}\|\nabla f(\x_k)-\g_k\|^2+ 4t^2L^2\mathbb{E}_{\mu_0}\|\nabla f(\x_k)\|^2+4tdL^2.
\end{align}
The third term can be bounded as follows: We choose an optimal coupling $\x_k \sim \mu_0(\x_k)$ and $\x^{\star} \sim \pi$ so that $\mathbb{E}\|\x_k-\x^{\star}\| = \W_2(\mu_0, \pi)^2$, then using Young's inequality and smoothness of $f$,
\begin{align}
    \label{eq:expectation-norm}
    \mathbb{E}_{\mu_0}\| \nabla f (\x_k)\|^2 &\leq 2   \mathbb{E}_{\mu_0}\|\nabla f(\x_k) - \nabla f(\x^{\star}) \|^2 +2    \mathbb{E}_{\mu_0}\|\nabla f(\x^{\star})\|^2\\
    &\leq 2L^2    \mathbb{E}_{\mu_0}\|\x_k - \x_0\|^2 + 2   \mathbb{E}_{\mu_0} \|\nabla f(\x^{\star})\|^2\\
    &\leq 2L^2  \W_2(\mu_0, \pi)^2 + 2dL\\
    &\leq \frac{4L^2}{\alpha}\KL(\mu_0||\pi) + 2dL.
\end{align}
The last inequality follows from Talagrand's inequality. Hence for  $t\leq \eta$ and $\eta\leq \frac{1}{2L}$, we have
\begin{align}
      \frac{d}{dt}\KL(\mu_t||\pi)&\leq -\frac{3}{4}\FI(\mu_t||\pi)+ (4t^2L^2 + 2)\mathbb{E}_{\mu_0}\|\nabla f(\x_k)-\g_k\|^2+ \frac{16t^2L^4}{\alpha}\KL(\mu_0||\pi) +4tdL^2+8t^2dL^3\\
      &\leq  -\frac{3\alpha }{2}\KL(\mu_t||\pi)+ (4t^2L^2 + 2)\mathbb{E}_{\mu_0}\|\nabla f(\x_k)-\g_k\|^2+ \frac{16t^2L^4}{\alpha}\KL(\mu_0||\pi) +4tdL^2+8t^2dL^3\\
      &\leq  -\frac{3\alpha }{2}\KL(\mu_t||\pi)+3\mathbb{E}_{\mu_0}\|\nabla f(\x_k)-\g_k\|^2+ \frac{16\eta^2L^4}{\alpha}\KL(\mu_0||\pi) +8\eta dL^2\\
       &\leq  -\frac{3\alpha }{2}\KL(\mu_t||\pi)+3\sigma_{k}^2+ \frac{16\eta^2L^4}{\alpha}\KL(\mu_0||\pi) +8\eta dL^2.
\end{align}
The second inequality is due to~\cref{eq:LSI}. Equivalently, we can write,
\begin{align}
     \frac{d}{dt}(e^{3\alpha t/2}\KL(\mu_t||\pi))\leq e^{3\alpha t/2}\left( 3\sigma_{k}^2+ \frac{16\eta^2L^4}{\alpha}\KL(\mu_0||\pi) +8\eta dL^2\right).
\end{align}
Integrating from $t=0$ to $t=\eta$ gives,
\begin{align}
    e^{3\alpha \eta/2}\KL(\mu_{\eta}||\pi) - \KL(\mu_0||\pi)\leq 6\eta \sigma_{k}^2+\frac{32\eta^3L^4}{\alpha}\KL(\mu_0||\pi) +16\eta^2 dL^2
\end{align}
for $\eta \leq \frac{2}{3\alpha}$. Rearranging the terms,
\begin{align}
  \KL(\mu_{\eta}||\pi) \leq   e^{-3\alpha \eta/2} \left[\left(1+\frac{32\eta^3L^4}{\alpha}\right)\KL(\mu_0||\pi)+6\eta \sigma_{k}^2 +16\eta^2 dL^2\right].
\end{align}
Renaming $\mu_0 = \mu_k$ and $\mu_{\eta} = \mu_{k+1}$, we obtain the result in the statement.

\end{proof}
The statement in \cref{lem:LSI-one-step-convergence} is generic and can be applied to any LMC algorithm with stochastic gradients with bounded variance on the trajectory of the algorithm. Note that this is different from assuming that the variance is uniformly upper bounded. Instead, we set inner loop and variance reduction parameters so that the variance does not explode along the trajectory of the algorithm.

\section{Technical Proofs for Gradient Estimation}
\label{sec:gradient-estimation-appendix}
\phaseoracle*
\begin{proof}
The proof constructs a sequence of unitary operators using the binary-to-phase conversion algorithm for different quantiles of \( X \). We begin by randomly drawing a classical sample \( s \) from the distribution that generates \( X \). By Chebyshev's inequality,
\begin{align}
   \Pr[|s-\mathbb{E}[X]]|\geq 3\sigma]\leq \frac{1}{9}. 
\end{align}
We consider the case $|s-\mathbb{E}[X]|$ is smaller than $3\sigma$ which holds with probability $8/9$. Next, we define the random variable \( Y = X - s \). Additionally, we introduce a random variable \( Y_{a,b} \), a truncated version of \( Y \), where values of \( Y \) outside the interval \( [a, b) \) are set to zero. The expectation \( \mathbb{E}[Y_{0,\infty}] \) can be expressed as a sum:
\begin{align}
   \mathbb{E}[Y_{0,\infty}] = \mathbb{E}[Y_{0,1}] + \sum_{k=1}^K 2^k \mathbb{E}\left[\frac{Y_{2^{k-1}, 2^k}}{2^k}\right] + \mathbb{E}[Y_{2^K, \infty}]. 
\end{align}
We define the unitary operator \( P^{Y_{a,b}}_{t, \epsilon} \), which implements the phase oracle for \(\mathbb{E}[Y_{a,b}] \) with an error of at most \( \epsilon \). The unitary \( P^{Y_{0,\infty}}_{t,\epsilon/2} \) can be implemented as the following product:
\begin{align}
  P^{Y_{0,\infty}}_{t,\epsilon/2} = P^{Y_{0,1}}_{t,\epsilon/6} \left( \prod_{k=1}^K P^{Y_{2^{k-1},2^k}}_{t,\epsilon/6K} \right) P^{Y_{2^K,\infty}}_{t,\epsilon/6}.  
\end{align}
When \( K = \log\left(\frac{120\sigma^2 t}{\epsilon}\right) \), the operator \( P^{Y_{2^K,\infty}}_{t,\epsilon/6} \) is effectively the identity operator, as:
\begin{align}
    \left| \ket{0} - e^{it \mathbb{E}[Y_{2^K,\infty}]} \ket{0} \right| \leq t \mathbb{E}[Y_{2^K,\infty}] \leq \frac{\epsilon}{6}.
\end{align}
The last inequality holds because:
\begin{align}
 \mathbb{E}[Y_{2^K,\infty}] &= \sum_{Y \geq 2^K} \Pr(Y) Y \leq \sum_Y \frac{1}{2^K} \Pr(Y) Y^2 = \frac{\mathbb{E}\|Y\|^2}{2^K}\\ 
 &\leq \frac{ 2 \mathbb{E}\|X - \mathbb{E}[X]\|^2 + 2\|s - \mathbb{E}[X]\|^2}{2^K} \\
 &\leq \frac{20\sigma^2}{2^K}= \frac{\epsilon}{6t},
\end{align}
where the inequality in the second line follows from the definition of $Y$ and Young's inequality. Since \( X_{0,1} \) is bounded between 0 and 1, we can implement \( P_{t, \epsilon/6}^{Y_{0,1}} \) using \( \Tilde{\Ocal}(1) \) queries to \( X \) via the binary-to-phase conversion algorithm (Lemma 2.12 in \cite{Cornelissen_2022}). We need to show how to implement \( P_{t, \epsilon/6K}^{Y_{a,b}} \) when \( b > 1 \). We start by defining the unitary operator:
\begin{align}
    V_{a,b} : \ket{0}\ket{0} &\mapsto \sum_Y \sqrt{\Pr(Y)} \ket{Y_{a,b}/b} \ket{0}\\
    &\mapsto \sum_Y \sqrt{\Pr(Y)} \ket{Y_{a,b}/b} \left( \sqrt{Y_{a,b}/b} \ket{0} + \sqrt{1 - Y_{a,b}/b} \ket{1} \right)\\
    &=\sqrt{\mathbb{E}[Y_{a,b}/b]}  \ket{\psi_0}\ket{0} + \sqrt{1 - \mathbb{E}[Y_{a,b}/b]} \ket{\psi_1} \ket{1}, 
\end{align}
where the $\ket{\psi_0}$ and $\ket{\psi_1}$ are normalized quantum states. 
Noting that
\begin{align}
  \mathbb{E}[Y_{a,b}/b] &\leq \frac{1}{b} \sum_{a\leq Y\leq b} \Pr(Y)Y\leq \frac{1}{ab} \sum_{a\leq Y\leq b} \Pr(Y)Y^2\\ 
  &= \frac{1}{ab} \mathbb{E}\|Y\|^2 \leq \frac{\sigma^2}{ab},  
\end{align}
we can apply the linear amplitude amplification algorithm (see \cite[Proposition 2.10]{Cornelissen_2022}) to implement the operator:
\begin{align}
   W_{a,b} : \ket{0}\ket{0} \mapsto \sqrt{p_{a,b}} \ket{\psi_0} \ket{0} + \sqrt{1 - p_{a,b}} \ket{\psi_1} \ket{1}, 
\end{align}
such that,
\begin{align}
   \left| \sqrt{p_{a,b}} - \sqrt{\frac{\mathbb{E}[Y_{a,b}/b]}{\sigma^2/(ab)}} \right| \leq \frac{\epsilon}{24Ktb} 
\end{align}
using \( \Tilde{\Ocal}(\sqrt{ab}/\sigma) \) calls to \( V_{a,b} \).
Let $t' = t\sigma^2/a$. Using the binary-to-phase conversion algorithm, we then implement \( \ket{\phi_{a,b}} = e^{it\mathbb{E}[Y_{a,b}] }\ket{0} \) with \( \Tilde{\Ocal}(t \sigma^2 / a) \) calls to \( W_{a,b} \) up to an operator norm error of at most \( \frac{\epsilon}{12K} \). By using the triangular inequality,
\begin{align}
    \|W_{a,b}\ket{0}-e^{it\mathbb{E}[Y_{a, b}] }\ket{0}\|&=  \|e^{it'p_{a,b}}\ket{0}-e^{it\mathbb{E}[Y_{a, b}] }\ket{0}\|\\
    &\leq t' \left| p_{a,b} - \frac{\mathbb{E}[Y_{a,b}/b]}{\sigma^2/(ab)} \right| + \frac{\epsilon}{12K}\\
    &\leq 2t' \left| \sqrt{p_{a,b}} - \sqrt{\frac{\mathbb{E}[Y_{a,b}/b]}{\sigma^2/(ab)}} \right| + \frac{\epsilon}{12K}\\
    &\leq \frac{\epsilon}{6K}.
\end{align}
Thus, the total implementation of \( P_{t, \epsilon/6K}^{Y_{a,b}} \) requires \( \Tilde{\Ocal}(t\sigma \sqrt{a/b}) \) calls to \( V_{a,b} \). This implies that each term in the product can be implemented using \( \Tilde{\Ocal}(t\sigma) \) quantum experiments and binary query oracles to \( Y \). Finally, we apply the phase \( e^{its} \) to the resulting state to implement \( P_{t,\epsilon/2}^{X_{0,\infty}} \). Similarly, we use the same method to implement \( P_{t,\epsilon/2}^{X_{-\infty,0}} \), and take the product:
\begin{align}
  P_{t,\epsilon}^{X} = P_{t,\epsilon/2}^{X_{0,\infty}} P_{t,\epsilon/2}^{X_{-\infty,0}} .  
\end{align}
This concludes the proof.
\end{proof}

\begin{lemma}
\label{lem:zeroth-order-gradient1}
    Suppose we run \cref{algo:grad_est} with the phase oracle in \cref{prop:phase-oracle} with evaluation accuracy $\epsilon' = \frac{\epsilon^2}{d^2\beta}$ to $f(\x,\xi)$. Let $\Tilde{\g}$ denote the output. Then, under \cref{ass:full-smoothness,ass:bounded-variance-zeroth-order},
    \[
    \|\Tilde{\g} - \nabla f(\x)\|\leq \epsilon,
    \]
    with probability at least $5/9$ using $\Tilde{\Ocal}(\frac{\sigma d}{\epsilon})$ queries to $f(\x;\xi)$. 
\end{lemma}

\begin{proof}
To be able to run the quantum gradient estimation algorithm, we need to implement $O_F$ that maps
\begin{equation}
    O_F\ket{\x}\mapsto e^{i \mathbb{E}_{\xi}[F(\x, \xi)] }\ket{\x},
\end{equation}
where $F(\x; \xi) = \frac{N}{2L l} (f(\x_0+\frac{l}{N}(\x-N/2);\xi)-f(\x_0;\xi))$. Let $\y=\frac{l}{N}(\x-N/2)$, the variance of $F(\x, \xi)$ is
\begin{align}
   \mathbb{E}\|F(\x;\xi)-\mathbb{E}[F(\x;\xi)]\|^2 &= \mathbb{E}\left\|\int_{0}^1\langle \nabla f(\x+t\y;\xi) -\nabla f(\x+t\y), \y \rangle dt  \right\|^2 \\
   &\leq  \|y\|^2\int_{0}^1\mathbb{E}\|\nabla f(\x+t\y;\xi) -\nabla f(\x+t\y)\|^2 dt  \\
   &\leq \sigma^2l^2d.
   \end{align}
Hence, implementing $e^{i \mathbb{E}[F(\x,\xi)] }$ takes $\Tilde{\Ocal}(\sigma l d^{1/2} \frac{N}{Ll} )=\Tilde{\Ocal}(\frac{\sigma}{\epsilon'^{1/2}\beta^{1/2}} )= \Tilde{\Ocal}(\frac{\sigma d}{\epsilon})$ queries to stochastic zeroth-order oracle and succeeds with probability $8/9$. Since \cref{algo:grad_est} uses $\Tilde{\Ocal}(1)$ queries to $O_F$ by \cref{lem:jordan_whp} and succeeds with probability $2/3$, the total query complexity is $\tilde{\Ocal}(\frac{\sigma d}{\epsilon})$ and success probability is at least $5/9$ due to union bound.
\end{proof}

\highprobgradient*
\begin{proof}
As the algorithm essentially computes the expectation of $\mathbb{E}_{\xi}[\tilde{\g}(\x,\xi)]$, we need to prove that $\mathbb{E}_{\xi}[\tilde{\g}(\x,\xi)]$ is  close to $\nabla f(\x)$. We consider the case that \cref{algo:grad_est} returns $\epsilon/8$ accurate estimate whenever the function $f$ behaves like $\beta$ smooth inside the grid points. Furthermore, we consider the case $\|\mathbf{s}-\nabla f(\x)\|\leq 2\sigma$. Both conditions are in fact achieved with high probability. Let $S\subseteq \Xi $ be a set such that the output of quantum gradient estimation (\cref{algo:grad_est}) $\g$ satisfies $\|\g-\nabla f(\x, \xi)\|\leq \frac{\epsilon}{8}$. Let $S' = \Xi - S$. We can consider the difference in $L_2$ norm separately for $S$ and $S'$ using the triangle inequality.
\begin{align}
    \|\mathbb{E}_{\xi} \tilde{\g}(\x, \xi) - \nabla f(\x) \|   &\leq  \|\mathbb{E}_{S} (\tilde{\g}(\x, \xi) - \nabla f(\x;\xi)) \| +  \|\mathbb{E}_{S'} (\tilde{\g}(\x, \xi) - \nabla f(\x;\xi)) \|.  
\end{align}
We first analyze the first term. The contribution to the first term is either due to gradient estimation error $\frac{\epsilon}{8}$ or it is due to the fact that $\g$ is replaced by $\mathbf{s}$ because $\|\g - \mathbf{s}\|>D$. Suppose that $S_1 = \{\xi\in \Xi: \|\g(\x;\xi)-s\|\leq D \}$ and $S_2= S-S_1$. We can separate the error further for both cases using the triangle inequality.

\begin{align}
\|\mathbb{E}_{S} (\tilde{\g}(\x, \xi)- \nabla f(\x;\xi)) \| &\leq \mathbb{E}_{\xi\in S_1} \|(\g(\x, \xi)- \nabla f(\x, \xi))\| + \mathbb{E}_{\xi \in S_2}\| (\mathbf{s}- \nabla f(\x, \xi))\|\\
&\leq \mathbb{E}_{\xi\in S_1} \|(\g(\x, \xi)- \nabla f(\x, \xi))\| + \mathbb{E}_{\xi \in S_2} \|(\mathbf{s}- \g(\x;\xi)\|\\& +\mathbb{E}_{\xi \in S_2} \|(\g(\x;\xi)- \nabla f(\x, \xi))\|\\
&\leq \frac{\epsilon}{8}+ \frac{\mathbb{E} \|\mathbf{s}- \nabla f(\x, \xi))\|^2}{D} + \frac{\epsilon}{8}.
\end{align}
The first inequality is due to the fact that for any $\xi \in S_2$, \cref{algo:QuantumStochasticGradient} replaces $\g$ by $\mathbf{s}$. 
The last inequality follows from the fact that $\|\g(\x;\xi)-\nabla f(\x;\xi)\|\leq\frac{\epsilon}{8}$ for any $\xi \in S$ and  $\mathbb{E}_{\xi \in S_2}\|\mathbf{s}-\g(\x;\xi)\|\leq \frac{\mathbb{E}\|\mathbf{s}-\g(\x;\xi)\|^2}{D}$ since for any $\xi \in S_2$ we have $\|\g(\x;\xi)-\mathbf{s}\|>D$. As $\|\mathbf{s}-\nabla f(\x)\|\leq 2\sigma$,
\begin{align}
    \mathbb{E}\|\mathbf{s}-\g(\x;\xi)\|^2&\leq  2\mathbb{E}\|\mathbf{s}-\nabla f(\x;\xi)\|^2+ 2 \mathbb{E}\|\nabla f(\x;\xi)-\g(\x;\xi)\|^2\\
    &\leq 10 \sigma^2.
\end{align}
Then, for $D = \frac{40\sigma^2}{\epsilon}$, we have $\frac{\mathbb{E}\|\mathbf{s}-\g(\x;\xi)\|^2}{D}\leq \frac{\epsilon}{4}$. Therefore, $ \|\mathbb{E}_{S} (\tilde{\g}(\x, \xi)- \nabla f(\x;\xi)) \|\leq \frac{\epsilon}{2}$.

The term due to $S'$ comes from the case where gradient estimation fails. Notice that whenever gradient estimation fails, we have $\|\tilde{\g}(\x;\xi) - \nabla f(\x)\| \leq \max(D, 2\sigma)$. Gradient estimation only fails when $f(\x;\xi)$ has a smoothness constant larger than $\beta$. Using Markov's inequality, this happens with probability at most $\frac{L}{\beta}$. Then,
\begin{align}
\mathbb{E}_{S'} \|(\tilde{\g}(\x, \xi)- \nabla f(\x)) \| \leq \frac{L}{\beta}\max(D, 2\sigma)\leq \frac{\epsilon}{4}
\end{align}
for $\beta  = \frac{160L\sigma^2}{\epsilon^2}$ and $\sigma\geq \epsilon$. This implies that non-smooth branches do not affect the expectation by replacing $\g$ with $\tilde{\g}$. Furthermore, the variance of $\tilde{\g}(\x)$ is
\begin{align}
    \mathbb{E}_{\xi}\| \tilde{\g}(\x, \xi) - \mathbb{E}[\tilde{\g}(\x, \xi)] \|^2 &\leq 2\mathbb{E}\left\|\tilde{\g}(\x,\xi) -\nabla f(x) \right\|^2+ 2\left\|\mathbb{E}[\tilde{\g}(\x,\xi)] -\nabla f(x) \right\|^2\\
    &\leq 2\mathbb{E}_{S'}\left\|\tilde{\g}(\x,\xi) -\nabla f(x) \right\|^2+2\mathbb{E}_{S}\left\|\tilde{\g}(\x,\xi) -\nabla f(x) \right\|^2+2\epsilon^2\\
    &\leq  \frac{2Ld}{\beta}\max(D^2, 4\sigma^2) + 2\mathbb{E}\left\|\nabla f(\x;\xi) -\nabla f(\x) \right\|^2 +2\mathbb{E}\left\|\mathbf{s}-\nabla f(\x) \right\|^2+3\epsilon^2\\
    &= \Ocal(\sigma^2).
\end{align}
Therefore we can use quantum mean estimation to output $\epsilon$ accurate vector $\boldsymbol{v}$ such that $\|\boldsymbol{v}-\nabla f(\x)\|\leq \epsilon$ using $\Tilde{\Ocal}(\sigma d^{1/2}/\epsilon)$ calls to algorithm $\mathcal{A}$. Since algorithm $\mathcal{A}$ uses $\Tilde{\Ocal}(1)$ queries to evaluation oracle, total stochastic evaluation complexity is $\Tilde{\Ocal}(\sigma d^{1/2}/\epsilon)$.
\end{proof}

\end{document}